\pdfoutput=1

\documentclass[]{aastex631}
\usepackage{amsmath}
\usepackage{graphicx}
\usepackage{mathrsfs}
\usepackage{color}
\usepackage{url}
\usepackage{CJKutf8}
\usepackage{ulem}
\usepackage{wasysym}
\usepackage{multirow}
\usepackage{booktabs}

\received{2022}
\revised{\today}
\accepted{2022}

\shorttitle{C/2018 F4}
\shortauthors{Hui et al. 2023}

\begin{document}

\title{
Splitting of Long-Period Comet C/2018 F4 (PANSTARRS)
}

\correspondingauthor{Man-To Hui}
\email{mthui@must.edu.mo}

\author{\begin{CJK}{UTF8}{bsmi}Man-To Hui (許文韜)\end{CJK}}
\affiliation{State Key Laboratory of Lunar and Planetary Science, 
Macau University of Science and Technology, 
Avenida Wai Long, Taipa, Macau}

\author{Michael S. P. Kelley}
\affiliation{Department of Astronomy, 
University of Maryland, College Park, MD 20742-0001, USA}

\author{Denise Hung}
\affiliation{Institute for Astronomy, University of Hawai`i, 
2680 Woodlawn Drive, Honolulu, HI 96822, USA}

\author{Tim Lister}
\affiliation{Las Cumbres Observatory, 
6740 Cortona Drive, Suite 102, Goleta, CA 93117, USA}

\author{Joseph Chatelain}
\affiliation{Las Cumbres Observatory, 
6740 Cortona Drive, Suite 102, Goleta, CA 93117, USA}

\author{Edward Gomez}
\affiliation{Las Cumbres Observatory, School of Physics and Astronomy, 
Cardiff University, Queens Buildings, The Parade, Cardiff CF24 3AA, UK}

\author{Sarah Greenstreet}
\affiliation{Department of Astronomy and the DIRAC Institute, 
University of Washington, 3910 15th Avenue NE, Seattle, WA 98195, USA}


\begin{abstract}

Long-period comet C/2018 F4 (PANSTARRS) was observed to show duplicity of its inner region in 2020 September, suggestive of a splitting event. We here present analyses of our observations of the comet taken from the LOOK project and the University of Hawaii 2.2 m telescope after the discovery of the splitting. The two fragments Components A and B, estimated to be $\sim\!60$ m to 4 km in radius, remained highly similar to each other in terms of brightness, colour, and dust morphology throughout our observing campaign from 2020 September to 2021 December. Our fragmentation model yielded that the two components split at a relative speed of $3.00 \pm 0.18$ m s$^{-1}$ in 2020 late April, implying a specific energy change of $\left(5.3 \pm 2.8 \right) \times 10^3$ J kg$^{-1}$, and that Component B was subjected to a stronger nongravitational acceleration than Component A in both the radial and normal directions of the orbit. The obtained splitting time is broadly consistent with the result from the dust morphology analysis, which further suggested that the dominant dust grains were millimeter-sized and ejected at speed $\sim\!2$ m s$^{-1}$. We postulate that the pre-split nucleus of the comet consisted of two lobes resembling the one of 67P, or that the comet used to be a binary system like main-belt comet 288P. Regardless, we highlight the possibility of using observations of split comets as a feasible manner to study the bilobate shape or binarity fraction of cometary nuclei.

\end{abstract}


\section{Introduction}
\label{sec_intro}

Fragmentation and disintegration are common fates for comets. Over the past 150 years, more than 40 comets have been observed to split \citep{2004come.book..301B}, and the number is still growing \citep[e.g.,][]{2009P&SS...57.1218F,2016ApJ...829L...8J,2015AJ....149..133L,2021AJ....162...70Y}. However, the mechanisms by which they fragment are usually far from clear, except in the case of D/1993 F2 (Shoemaker-Levy 9), which was torn apart by the tidal force of Jupiter \citep{1993Natur.365..733S,1994A&A...289..607S}. Other plausible splitting mechanisms include rotational instability, thermal stress, internal gas pressure, and impact \citep[and citations therein]{2004come.book..301B}. For long-period comets, their splitting may explain the fact that the observed number of returning members appears to be depleted in comparison to dynamically new counterparts, which is known as the fading problem of long-period comets \citep{2022AJ....164..158J,2002Sci...296.2212L,1950BAN....11...91O,1999Icar..137...84W}. Cometary splitting events tend to be rapidly evolving and short-lived in most cases, rendering difficulties in obtaining prompt and good-quality observations before observing windows vanish. Nevertheless, split comets are scientifically important in that they offer us precious opportunities to investigate their interiors, which are otherwise nearly impossible to probe via remote observations. Studies of split comets may even shed light on formation of the solar system, as the interiors of comets may reflect what material was available early on. \citep{2004come.book..301B}.

Recently, we sought an opportunity to closely monitor the splitting event at long-period comet C/2018 F4 (PANSTARRS). As the name suggests, the comet was initially discovered by the Pan-STARRS survey, on 2018 March 17 at a heliocentric distance of $r_{\rm H} = 6.4$ au. However, it was initially designated as an asteroidal object A/2018 F4 \citep{2018MPEC....F..139T}. With reports of cometary activity by follow-up observers, it was later reclassified as a comet \citep{2018MPEC....H...21S}. The orbital solution by JPL Horizons\footnote{\url{https://ssd.jpl.nasa.gov/tools/sbdb_lookup.html\#/?sstr=2018f4}, retrieved on 2022 December 29.} gives that the current osculating heliocentric orbit of the comet is slightly hyperbolic (eccentricity $e = 1.002$) and highly inclined with respect to the ecliptic (orbital inclination $i = 78\fdg1$), with a perihelion distance of $q = 3.4$ au, which the comet passed on TDB 2019 December 4. Despite the hyperbolic orbit, dynamical analyses show that the comet is from the Oort cloud rather than of extrasolar origin \citep{2019RNAAS...3..143D,2019A&A...625A.133L}. The comet appeared to be ordinary in the long-period comet population and showed no sign of hydrated altered minerals in spectroscopic observations \citep{2019A&A...625A.133L}. In 2020 September, the comet was reported by amateurs to have split into two pieces on its outbound leg.\footnote{\url{https://groups.io/g/comets-ml/message/29101}} To follow up the splitting, we immediately commenced collecting optical observations of the comet. In this paper, we present photometric and dynamical analyses of the nucleus duo of C/2018 F4 based on our observations, which are detailed in Section \ref{sec_obs}. Results of our photometric, morphological, and dynamical analyses are presented in Section \ref{sec_rslt}, followed by discussion in Section \ref{sec_disc} and a summary in Section \ref{sec_sum}.

\section{Observations}
\label{sec_obs}

\begin{figure*}
\begin{center}
\gridline{\fig{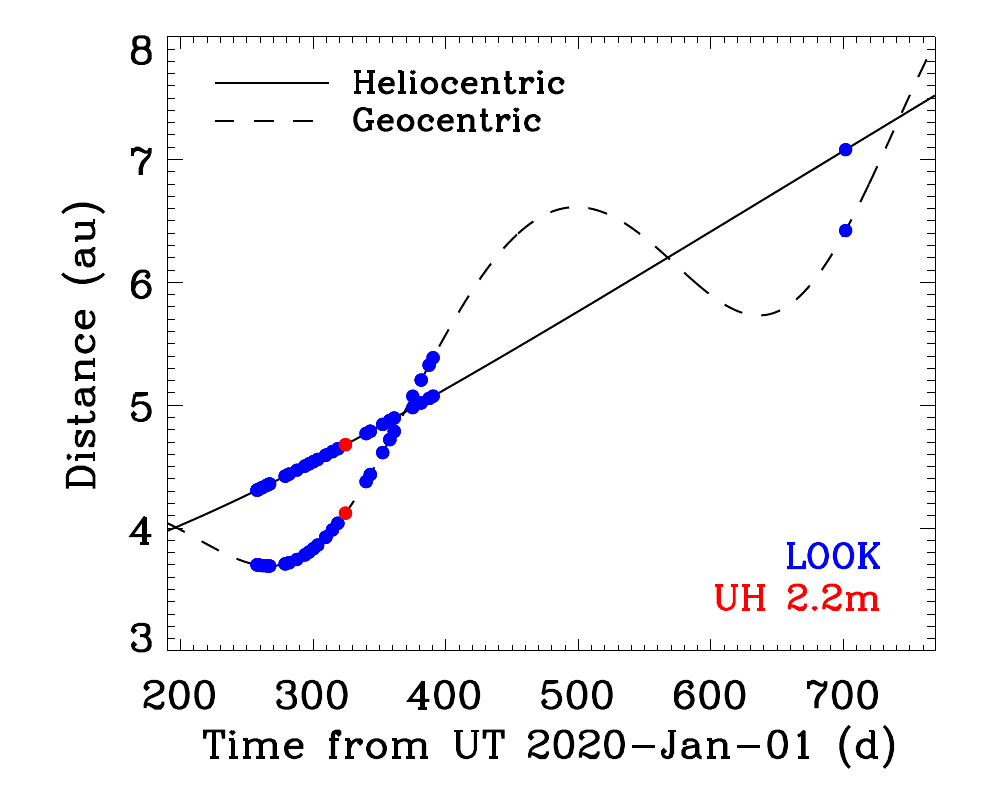}{0.5\textwidth}{(a)}
\fig{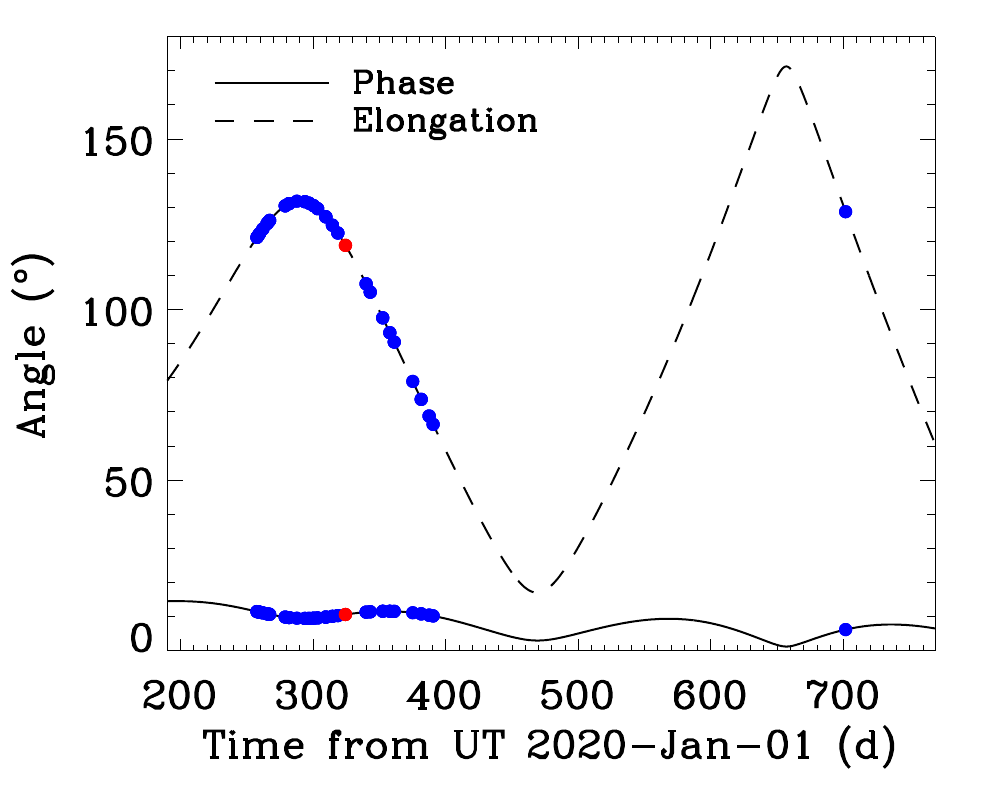}{0.5\textwidth}{(b)}}
\gridline{\fig{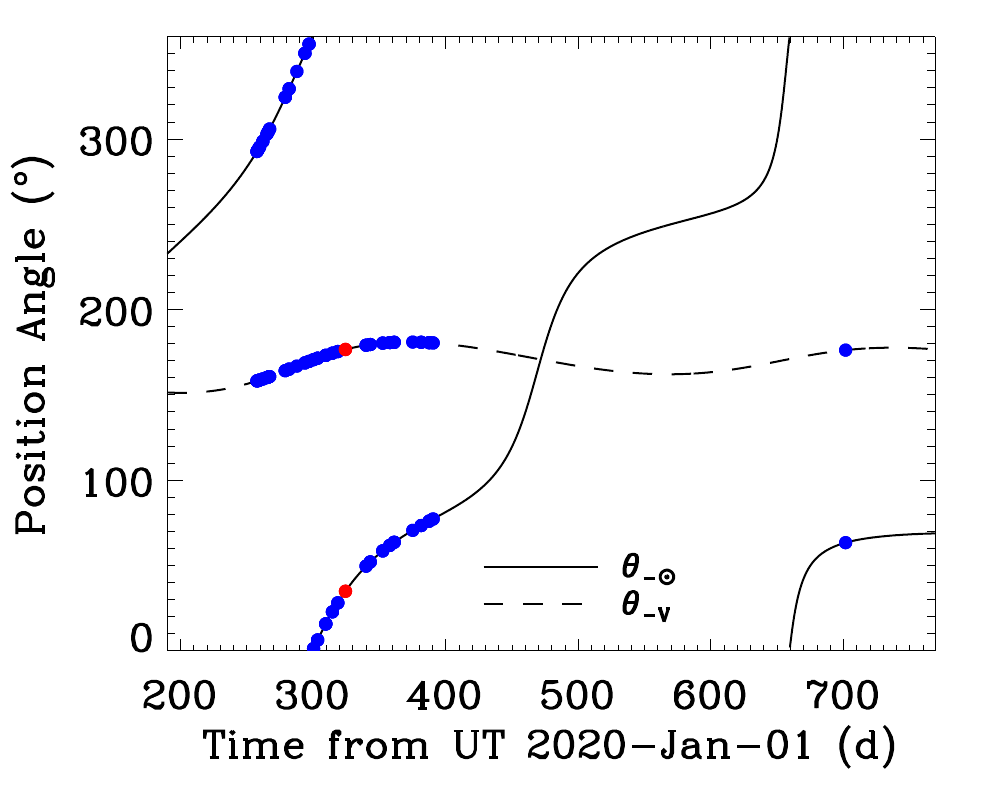}{0.5\textwidth}{(c)}
\fig{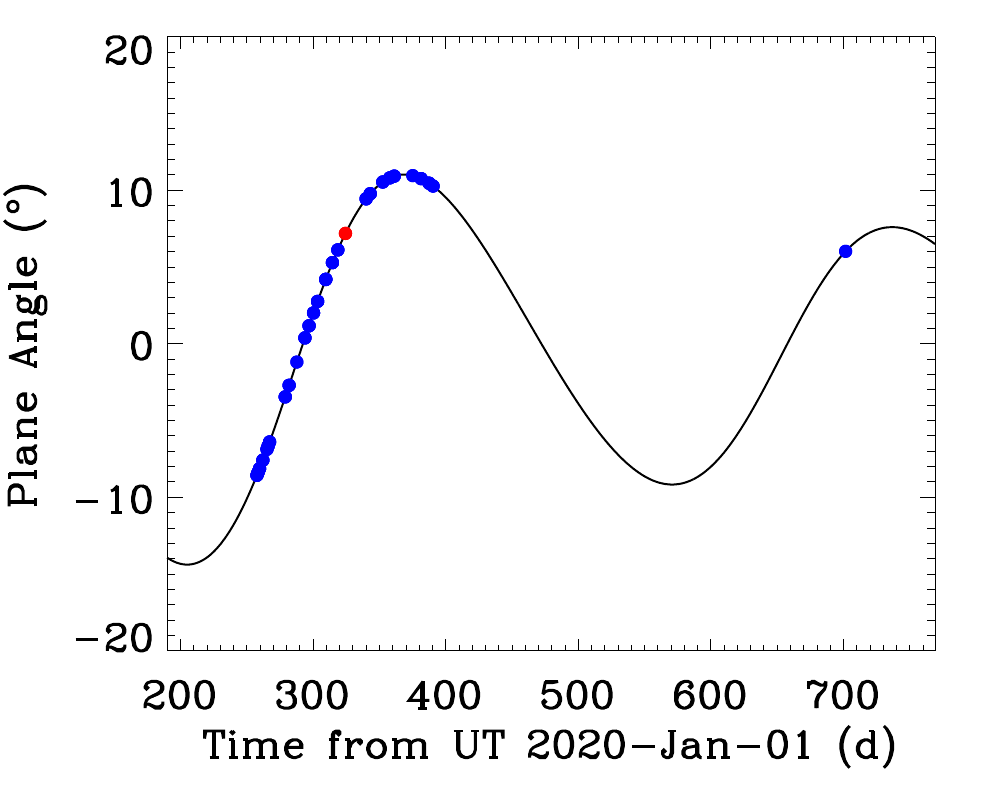}{0.5\textwidth}{(d)}}
\caption{
Observing geometry of C/2018 F4 varying with time during our observations from LOOK (blue) and UH 2.2 m (red). Here, we plot (a) heliocentric and geocentric distances, (b) phase and solar elongation, (c) position angles of antisolar direction ($\theta_{-\odot}$) and negative heliocentric velocity projected onto the sky plane ($\theta_{-{\bf V}}$), and (d) plane angle of the comet's orbit. 
\label{fig:vgeo}
} 
\end{center} 
\end{figure*}

\begin{figure*}
\epsscale{1.2}
\begin{center}
\plotone{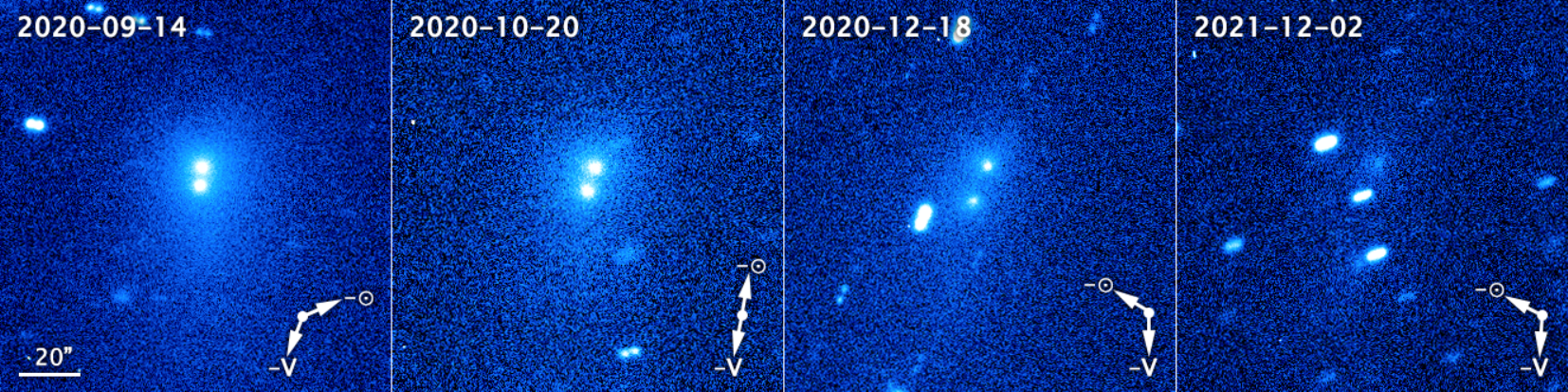}
\caption{
Composite LOOK images of comet C/2018 F4 in the $r^{\prime}$ band marked with observing dates in UTC. A scale bar of 20\arcsec~in length and position angles of the antisolar direction and negative heliocentric velocity projected onto the sky plane are marked. The images all have J2000 equatorial north pointing upwards and east pointing to the left. The comet had split into two major components by the earliest LOOK observation; we refer to the upper and lower ones as Components A and B (designated C/2018 F4-A and C/2018 F4-B by the Minor Planet Center), respectively.
\label{fig:obs_LOOK}
} 
\end{center} 
\end{figure*}

\begin{figure}
\epsscale{1.0}
\begin{center}
\plotone{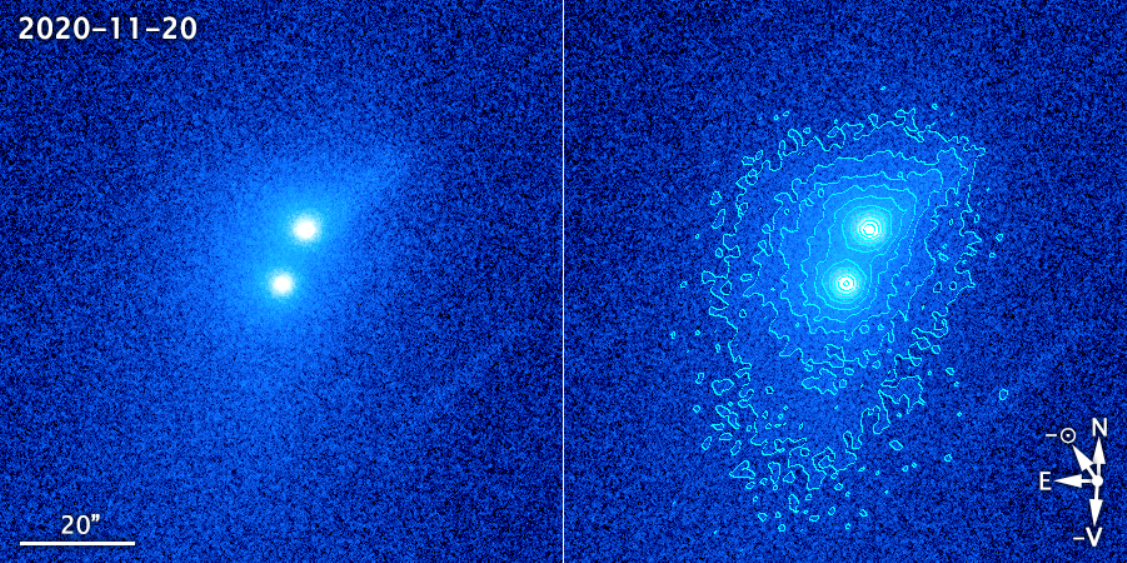}
\caption{
Median-combined $r^\prime$-band image of comet C/2018 F4 from the UH 2.2 m telescope on UTC 2020 November 20. The two panels show completely the same image except that the right one is overlaid with asinh-scale isophotal contours. As in Figure \ref{fig:obs_LOOK}, a 20\arcsec-long scale bar together with the comet's antisolar direction and on-sky component of the negative heliocentric velocity is shown. Also shown is a compass marking directions of J2000 equatorial north and east. The northwestward protrusion of Component A is due to a nearby background-star trail that was not thoroughly removed in the image-combining process. The smudge in the lower right corner of the image is from another lingering star trail.
\label{fig:obs_UH88}
} 
\end{center} 
\end{figure}

Our observations of C/2018 F4 were primarily obtained from the LCO Outbursting Objects Key (LOOK) Project \citep{2022PSJ.....3..173L} using seven identical 1 m telescopes at Siding Spring (Australia), Sutherland (South Africa), and Cerro Tololo (Chile), unevenly spanning from 2020 September to 2021 December. The images were taken through SDSS $g^\prime$ and $r^\prime$ filters and all have a square field of view (FOV) of $\sim \! 26\arcmin \times 26\arcmin$ with an angular resolution of 0\farcs39 pixel$^{-1}$. We measured seeing during these observations to be $\sim$1\arcsec-2\arcsec~FWHM. The acquired images were automatically processed by the LCO Beautiful Algorithms to Normalize Zillions of Astronomical Images \citep[BANZAI;][]{2018SPIE10707E..0KM} pipeline, which included bias and dark removal, flat-field correction, and astrometric solving in a realtime manner. In addition, we also obtained a successful single-night observation of the comet using the University of Hawaii (UH) 2.2 m telescope on the summit of Maunakea, Hawai`i, with a Tektronix $2048 \times 2048$ CCD camera at the f/10 Cassegrain focus, through {\it B}, {\it r}$^\prime$, and {\it i}$^\prime$ filters on 2020 November 20. These images have a square FOV of $7\farcm5 \times 7\farcm5$ and a pixel scale of 0\farcs22. During the observation, the seeing was $\sim \! 0\farcs8$ FWHM. We performed standard calibration for the images by subtracting bias and flatfielding, the latter of which was computed with additional science images in the same filters from the same night.

We plot the observing geometry of C/2018 F4 varying with time in Figure \ref{fig:vgeo}. All of our collected observations clearly show that comet C/2018 F4 exhibited a co-moving duplicity of its optocentre, both possessing their own comae (see selected composite images in Figures \ref{fig:obs_LOOK} and \ref{fig:obs_UH88}). This unambiguously confirms that the nucleus of the comet has split into two major components by the time of our observations.

\section{Results}
\label{sec_rslt}

\subsection{Photometry}
\label{ss_phot}

\begin{figure}
\epsscale{1.0}
\begin{center}
\plotone{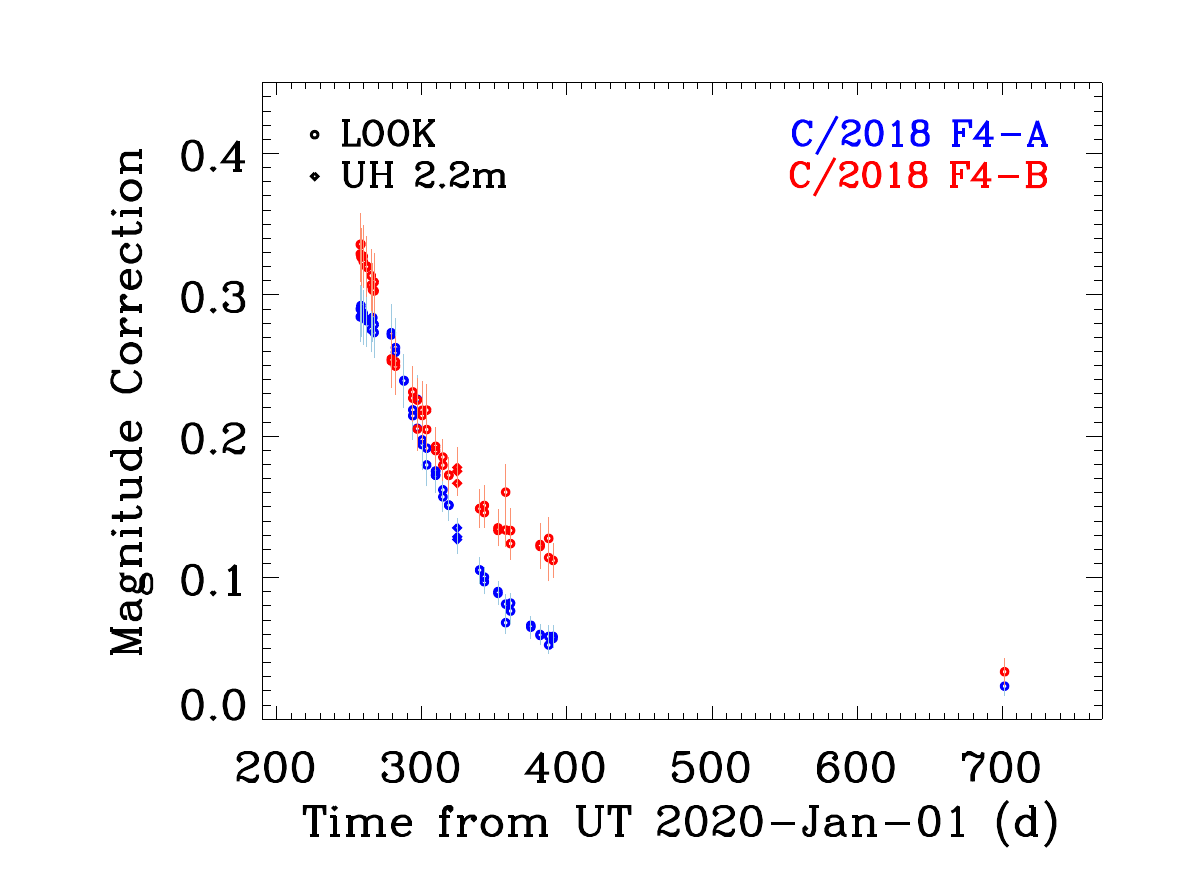}
\caption{
Magnitude corrections for the photometric measurements of C/2018 F4-A (blue) and B (red) with a circular aperture with a fixed radius of 8000 km projected at the distance of the comet as a function of time. Results from different facilities are discriminated by symbols shown in the legend. The gradually increasing apparent angular separation between the two components results in progressively smaller magnitude corrections. 
\label{fig:mcorr}
} 
\end{center} 
\end{figure}

Our images were photometrically calibrated using the methods described by \citet{2022PSJ.....3..173L} with {\tt calviacat} \citep{2019zndo...2635841K}. Briefly, individual frames were calibrated to the Pan-STARRS 1 (PS1) photometric system \citep{2012ApJ...750...99T} using the ATLAS Refcat2 catalog \citep{2018ApJ...867..105T} and background stars. The photometric calibration includes color corrections to match the PS1 filters. Photometry of each nuclear component of C/2018 F4 was measured with circular apertures having a fixed linear radius of 8000 km projected at the distance of the comet (corresponding to angular diameters varying from $\sim\!$3\farcs4~to 6\farcs0, depending on the observer-centric distance of the comet). The background was determined within sufficiently large adjacent annuli, the size of which depended on the extent of the split comet in each image so as to get rid of contamination therefrom as much as possible, but on the other hand, not overlarge lest contamination from background sources was encompassed (annular radii range from 20\arcsec~to 120\arcsec).

\begin{figure}
\epsscale{1.0}
\begin{center}
\plotone{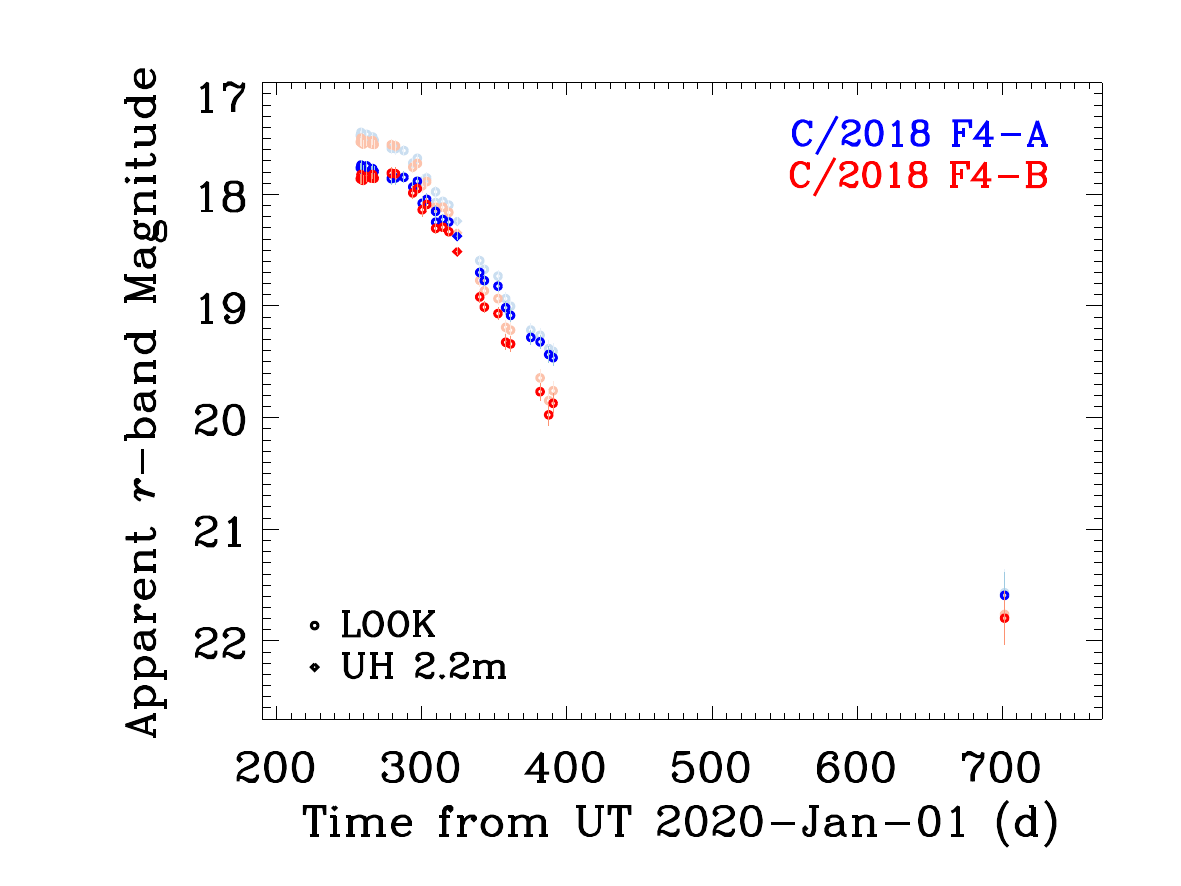}
\caption{
Apparent $r$-band magnitude of C/2018 F4-A and B measured with a circular aperture of fixed 8000 km radius projected at the distance of the comet versus time. Data points with and without the flux correction are plotted in darker and lighter colours, respectively.
\label{fig:lc}
} 
\end{center} 
\end{figure}

We then proceeded to correct the photometric measurements for each nuclear component so as to remove the mutual contamination from the partially overlapping comae, simplistically assuming that each coma is in steady state (see Appendix \ref{sec_apndx} for a detailed description of the derivation). In Figure \ref{fig:mcorr}, we plot the magnitude corrections for each nuclear component, both of which decrease with time as expected as the nuclear components apparently drifted further apart from each other, resulting in diminishing mutual contamination. We show the apparent $r$-band lightcurves of the two components after the magnitude corrections in Figure \ref{fig:lc}, in comparison to the ones without correcting for mutual contamination. In general, the correction did not alter the overall trend of each component's apparent magnitude lightcurve, which overall became fainter over the course of the observed timespan. At the beginning of our observing campaign, the apparent magnitude lightcurve of C/2018 F4-A was almost indistinguishable from the one of C/2018 F4-B, but thereafter the latter appeared to drop slightly more rapidly, making it systematically fainter than the other one (see Figure \ref{fig:lc}).

\begin{figure}
\epsscale{1.0}
\begin{center}
\plotone{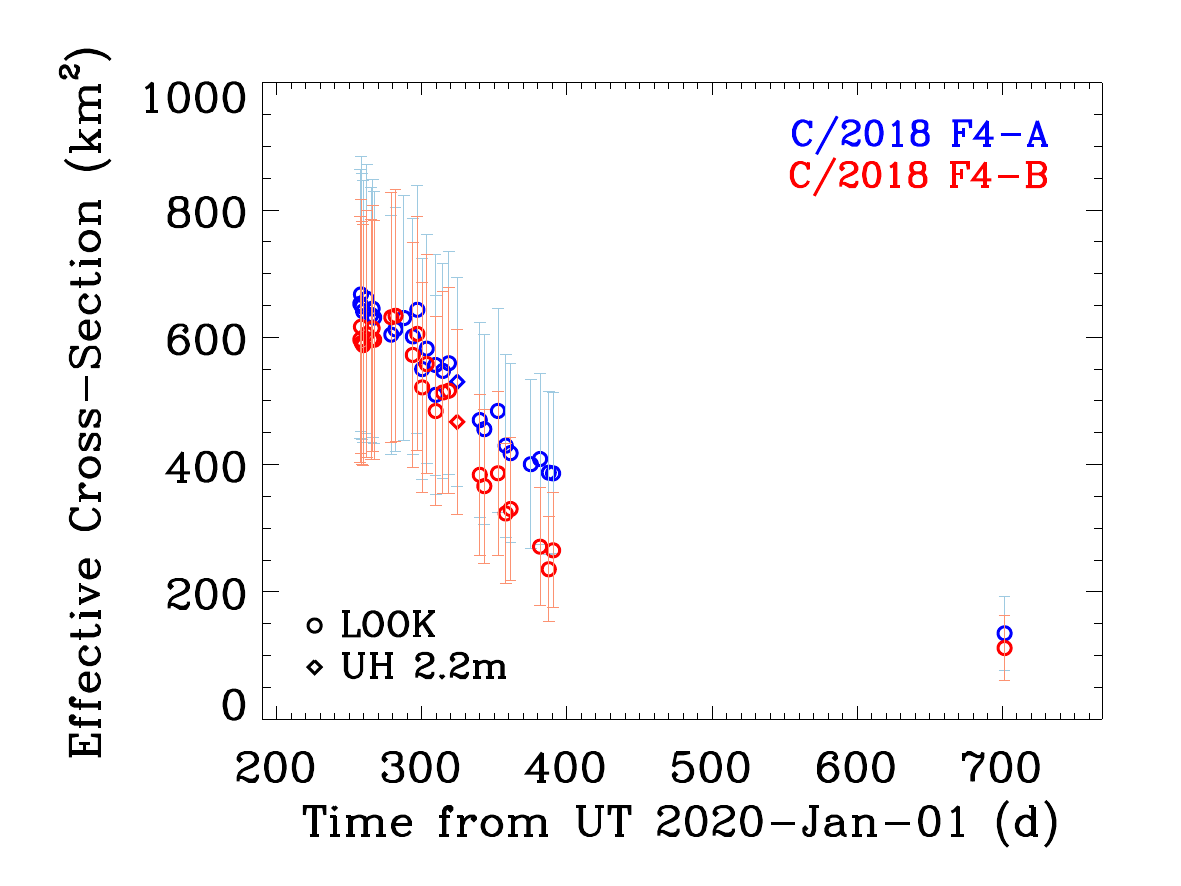}
\caption{
Temporal evolution of effective scattering cross-sections of C/2018 F4-A (blue) and B (red) within a circular aperture 8000 km in radius, assuming a constant {\it r}-band geometric albedo of $p_r = 0.05$.
\label{fig:xs}
} 
\end{center} 
\end{figure}

The apparent $r$-band magnitude of each nuclear component $m_{r}$ is straightforwardly related to its total effective scattering cross-section ${\it \Xi}_{\rm e}$ by 
\begin{equation}
{\it \Xi}_{\rm e} = \frac{\pi}{p_r \phi\left(\alpha\right)} \left(\frac{r_{\rm H} {\it \Delta}}{r_{\oplus}}\right)^{2} 10^{0.4 \left(m_{\odot,r} - m_{r}\right)}
\label{eq_XS},
\end{equation}
\noindent where $r_{\rm H}$ and $\it \Delta$ are the heliocentric and observer-centric distances, respectively, $r_{\oplus} = 1$ au is the mean Sun-Earth distance, $p_r$ is the $r$-band geometric albedo, $\phi\left(\alpha\right)$ is the phase function normalised at zero phase angle, and $m_{\odot,r} = -26.93$ is the apparent $r$-band magnitude of the Sun at $r_{\oplus} = 1$ au \citep{2018ApJS..236...47W}. We set $p_r = 0.05$ \citep{2018SSRv..214...64L} and adopted a linear phase model having a linear phase coefficient of $\beta_{\alpha} = 0.03 \pm 0.01$ mag deg$^{-1}$ \citep{1987A&A...187..585M}, both appropriate for comets. Here we did not attempt to incorporate the uncertainty in the geometric albedo, which cannot be constrained from our observations. Even if our assumed value is later found to be off, one can still easily scale and improve our estimates because of the inverse proportionality between the total effective scattering cross-section and geometric albedo. We show the total effective scattering cross-sections of both nuclear components varying with time in Figure \ref{fig:xs}, where the decline is obviously seen. We can think of two possibilities that may account for such trends -- 1) cometary activity of both components faded as they receded from the Sun, and 2) smaller dust grains were more efficiently swept away by solar radiation pressure and drifted beyond the photometric apertures than were the larger counterparts. Based upon our analysis of the dust morphology (see Section \ref{ss_morph}), the former is preferred. Assuming a bulk mass density of $\rho_{\rm d} \sim 1$ g cm$^{-3}$ for the dust grains, with the dominant dust grain size $\bar{\mathfrak{a}}_{\rm d} \sim 1$ mm (see Section \ref{ss_morph} also), we estimated that the corresponding average net mass-loss rate over the course of our observing campaign was $\left\langle \dot{M}_{\rm d} \right\rangle = 2\rho_{\rm d} \bar{\mathfrak{a}}_{\rm d} \left\langle \dot{{\it \Xi}}_{\rm e} \right\rangle /3 \approx 9 \pm 4$ kg s$^{-1}$ for both nuclear component.

\begin{figure*}
\begin{center}
\gridline{\fig{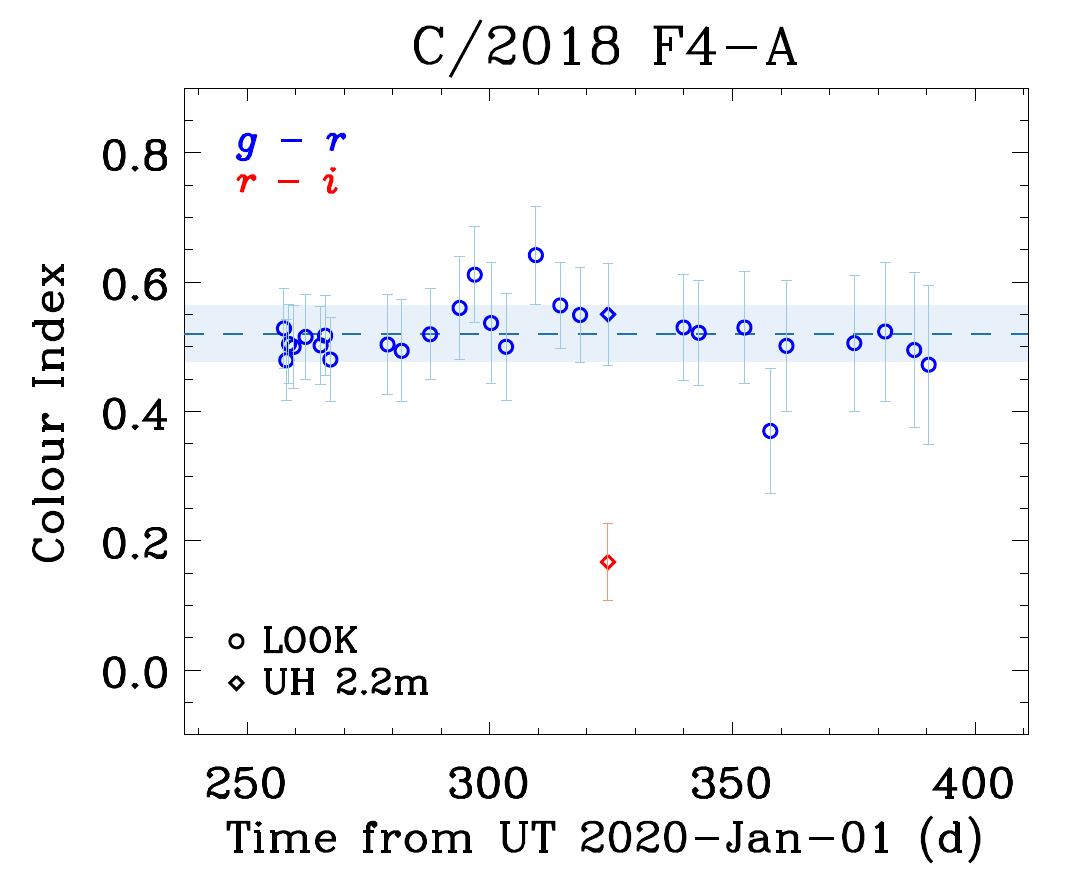}{0.5\textwidth}{(a)}
\fig{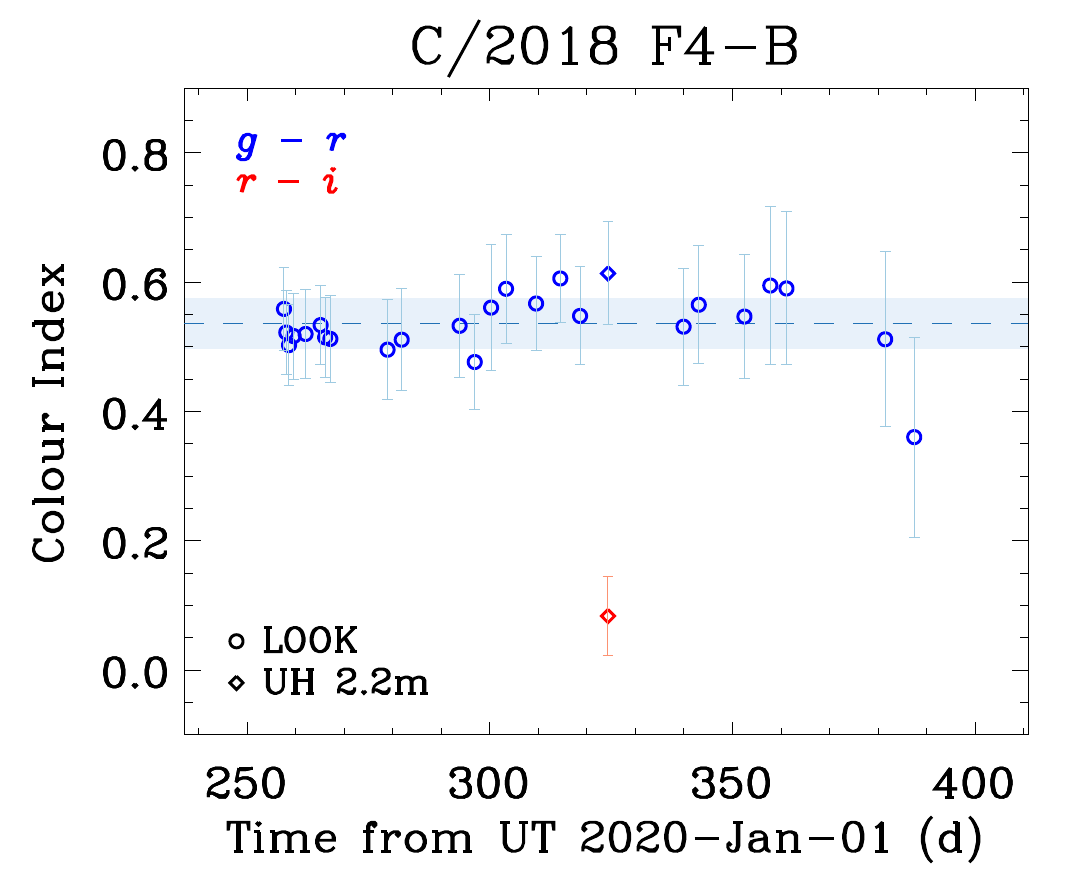}{0.5\textwidth}{(b)}}
\caption{
Colours of C/2018 F4-A (left) and B (right) in terms of $g-r$ and $r-i$ indices. The weighted mean values of the $g-r$ colours of the two components are shown as horizontal dashed lines in both panels, with the shaded regions representing the respective standard deviations as $\pm 1 \sigma$ uncertainties. No significant colour variation for either of the components beyond the noise level was seen. The two components shared a statistically similar colour.
\label{fig:clr}
} 
\end{center} 
\end{figure*}

Using our multiband observations, we examined the colours of the two nuclear components and the temporal trends thereof. Figure \ref{fig:clr} is a comparison between the colours of Component A and Component B in terms of their $g-r$ and $r-i$ colour indices. Unfortunately, our UH 2.2 m observation on 2020 November 20 was the only one in our campaign that covered the $i'$ band, and therefore we could only compare the $r-i$ colour indices of the two components from the single night. What we found is that in the $r-i$ regime, the mean colour of Component A, $\left\langle r-i\right\rangle_{\rm A} = 0.17 \pm 0.06$, is not statistically different than the counterpart of Component B, $\left\langle r-i\right\rangle_{\rm B} = 0.08 \pm 0.06$, where the reported errors are both standard deviations from repeated measurements. As for the $g-r$ regime, no statistically confident temporal colour trend is seen for either of the nuclear components (see Figure \ref{fig:clr}). We then calculated the weighted mean values of their $g-r$ colours over the course of our observing campaign to be $\left\langle g-r \right\rangle_{\rm A} = 0.52 \pm 0.04$ and $\left\langle g-r \right\rangle_{\rm B} = 0.54 \pm 0.04$ for Components A and B, respectively, which are again not statistically different. The errors thereof are standard deviations of the colour measurements. Strictly speaking, what we measured is most likely the colour from the dust environment rather than the nuclear components themselves, given the cometary appearances. Nevertheless, our measurements suggest that Components A and B had no major colour differences.

\citet{2019A&A...625A.133L} measured the normalised spectral slope of C/2018 F4 before the splitting event to be $S^{\prime} = \left(4.0 \pm 1.0 \right) \%$ per $10^3$ \AA~from their spectroscopic observations. To compare pre- and post-split colours of the comet, we had to also calculate the spectral slope from the colour measurements according to its definition by \citet{1984AJ.....89..579A} and \citet{1986ApJ...310..937J} as
\begin{equation}
    S^{\prime}\left(\lambda_{1}, \lambda_{2} \right) = -\left(\frac{2}{\Delta\lambda_{1,2}} \right) \frac{10^{0.4 \left[\Delta m_{1,2} - \Delta m_{1,2}^{\left(\odot\right)} \right]} - 1}{10^{0.4 \left[\Delta m_{1,2} - \Delta m_{1,2}^{\left(\odot\right)} \right]} + 1}
\label{eq_Sprime},
\end{equation}
\noindent where $\lambda_1$ and $\lambda_2$ are the wavelength range, $\Delta \lambda_{1,2} \equiv \lambda_1 - \lambda_2$ is the bandwidth, and $\Delta m_{1,2}$ and $\Delta m_{1,2}^{\left(\odot\right)}$ are the colour indices of the comet and the Sun, respectively, in the wavelength regime. The normalisation is chosen to be at the midpoint of the wavelength range. An object with $S^{\prime} = 0$ corresponds to its colour being completely the same as that of the Sun, while a negative value conveniently indicates a colour bluer than the solar colours, otherwise redder. Substituting with the $g-r$ and $r-i$ colour indices of the Sun and the corresponding bandwidths given in \citet{2018ApJS..236...47W}, Equation (\ref{eq_Sprime}) yields $S^{\prime} = \left(3.8 \pm 3.0 \right) \%$ per $10^3$ \AA~for Component A and $\left(4.8 \pm 2.8\right) \%$ per $10^3$ \AA~for Component B as their normalised spectral slopes in the $g-r$ regime, while their slopes in the $r-i$ regime are $\left(3.3 \pm 4.4\right) \%$ per $10^3$ \AA~and $\left(-2.5 \pm 4.5 \right) \%$ per $10^3$ \AA, respectively. Thus, the colours of Components A and B were both consistent with that of pre-split C/2018 F4 reported by \citet{2019A&A...625A.133L} within the noise level. This result, together with the similarity in the colours of the two components, may imply the homogeneity of the cometary nucleus of C/2018 F4. As a comparison, the two lobes of the nucleus of Jupiter-family comet 67P/Churyumov-Gerasimenko were also found to share a remarkably similar colour \citep{2015Sci...347a1044S}. The colour of the comet we obtained resembles those of many other long-period comets reported in \citet{2015AJ....150..201J}.

\subsection{Dust Morphology}
\label{ss_morph}

\begin{figure*}
\begin{center}
\gridline{\fig{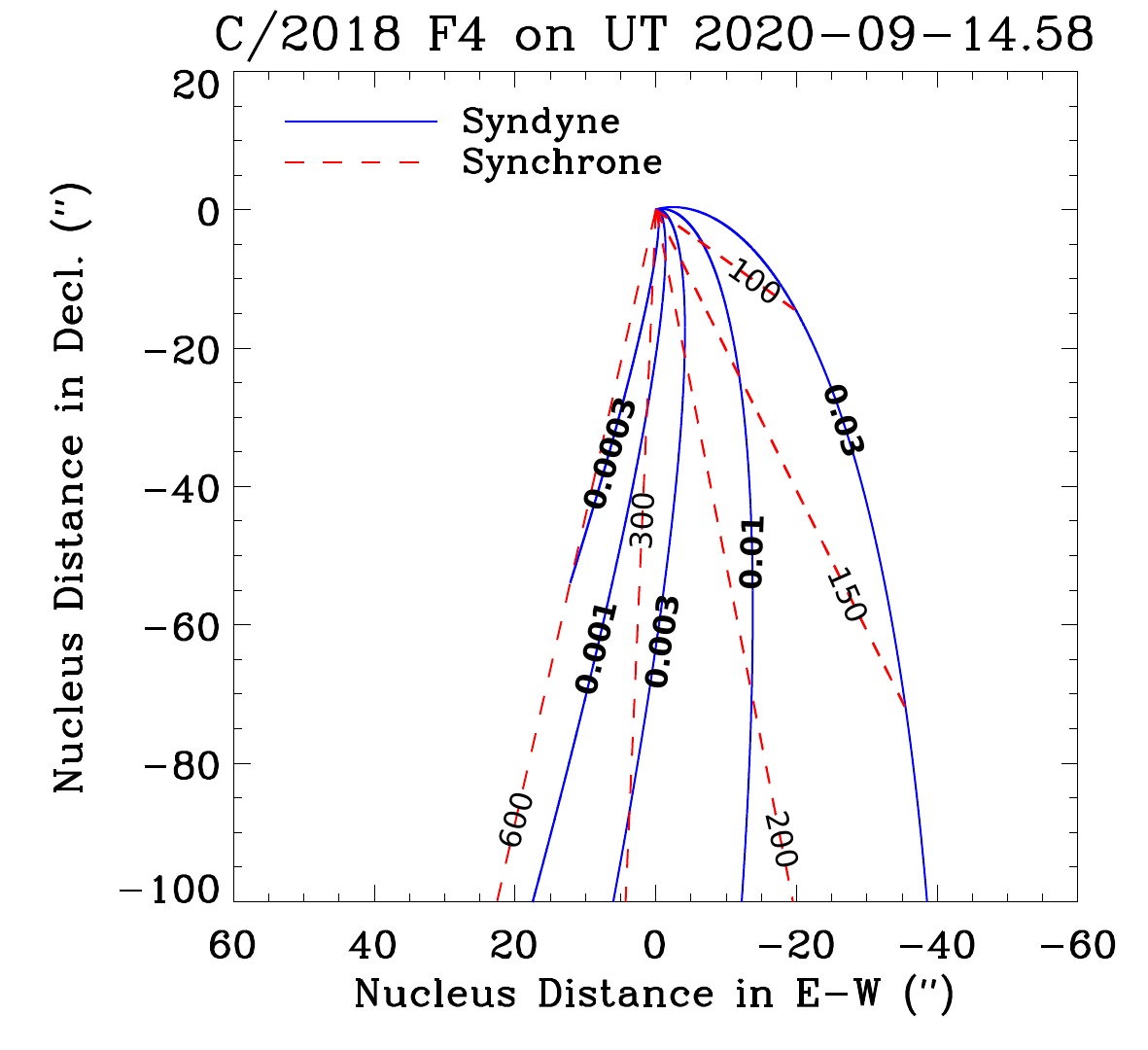}{0.5\textwidth}{(a)}
\fig{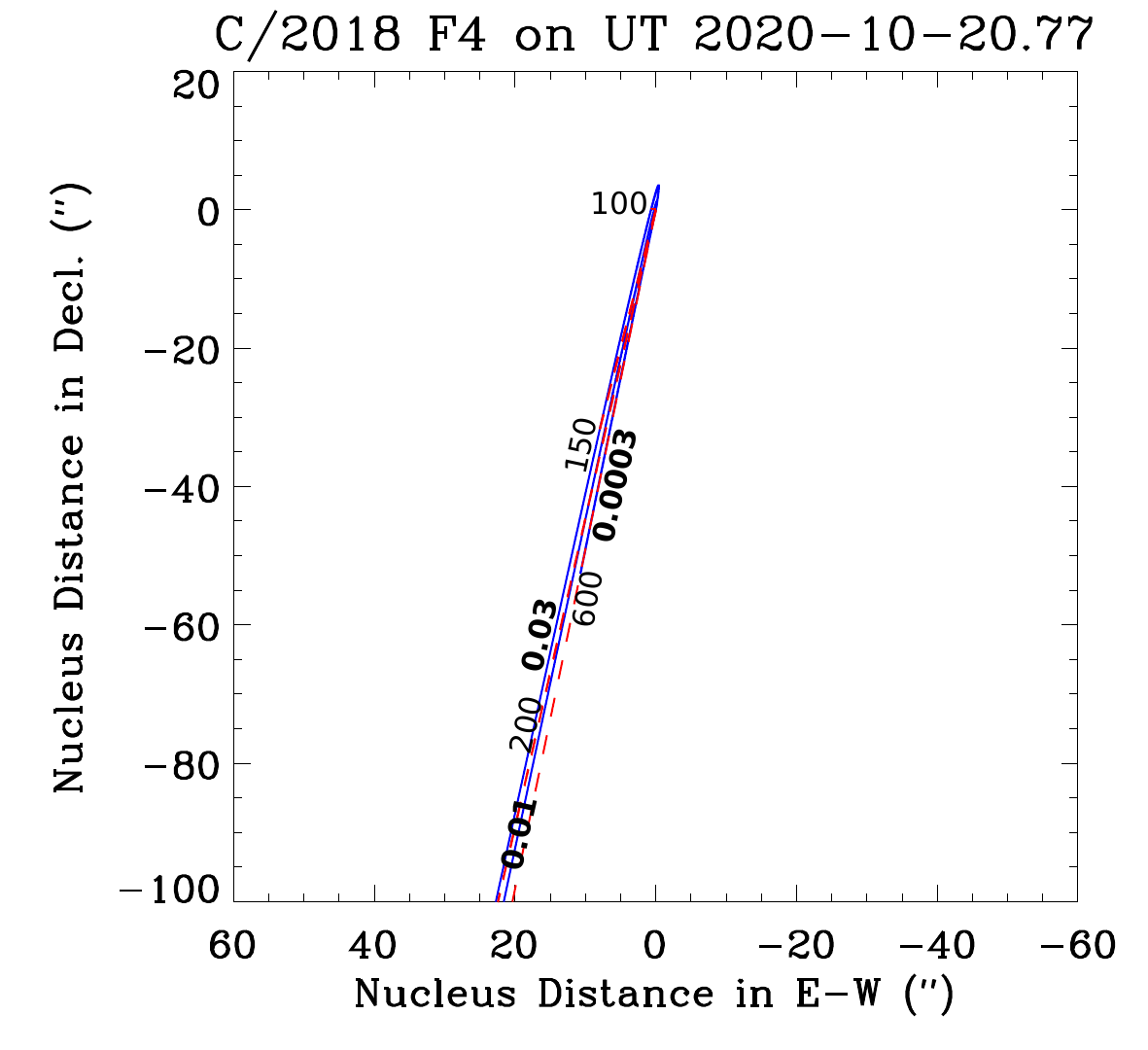}{0.5\textwidth}{(b)}}
\gridline{\fig{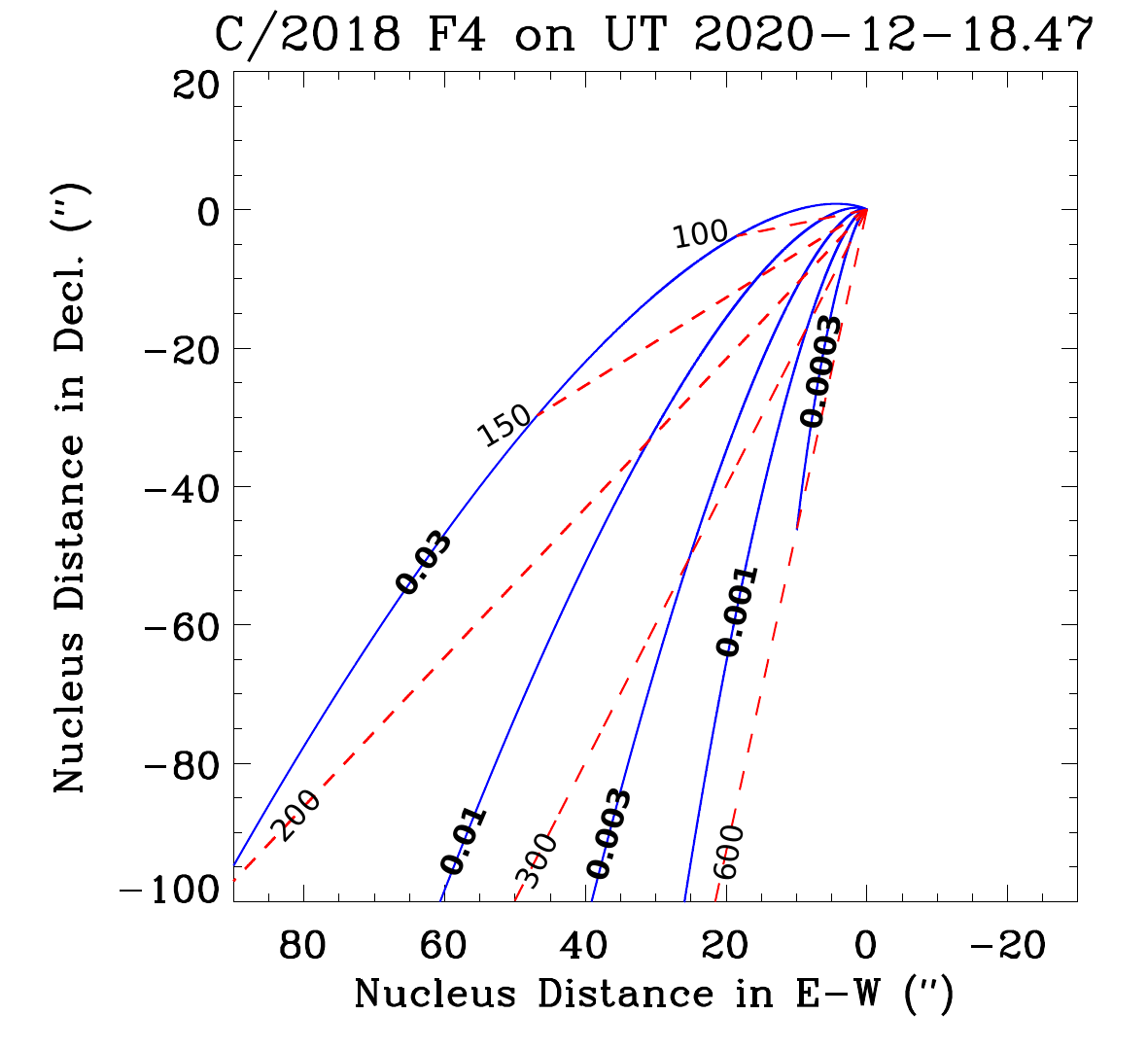}{0.5\textwidth}{(c)}
\fig{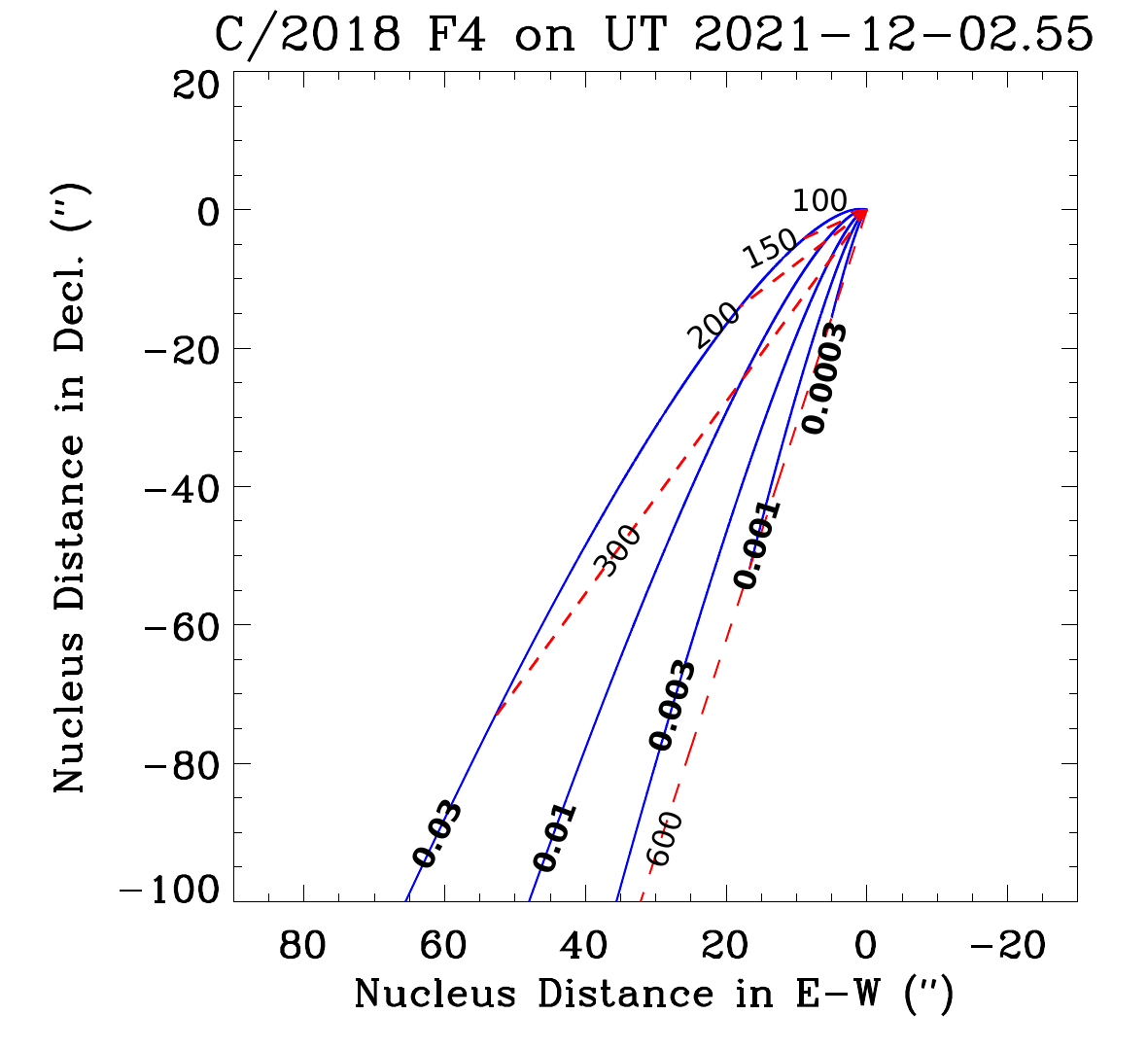}{0.5\textwidth}{(d)}}
\caption{
Computed syndyne-synchrone grids of C/2018 F4 corresponding to the selected LOOK observations in Figure \ref{fig:obs_LOOK}. In each panel, syndynes and synchrones are plotted as blue solid and red dashed lines, respectively. The values of the $\beta$ parameter of the plotted syndynes are labelled in bold weight, while the numbers in regular weight represent the dust release time from the observed epochs in days.
\label{fig:FP}
} 
\end{center} 
\end{figure*}

The dust morphology of comet C/2018 F4 implies the physical properties of its dust environment. Here, we applied the model by \citet{1968ApJ...154..327F} to study the dust morphology of the comet. Because the comet had split into two major fragments by the time of our earliest observation, each exhibiting a coma, we studied each component separately. In the Finson-Probstein model, all dust grains are assumed to leave the cometary nucleus at zero speed. The subsequent motion of the dust is then governed by parameter $\beta$, the ratio between accelerations of solar radiation pressure and solar gravitation, which is related to the grain size $\mathfrak{a}_{\rm d}$ and bulk density $\rho_{\rm d}$ by
\begin{equation}
    \beta = \frac{\mathcal{C}_{\rm pr} \mathcal{Q}_{\rm pr}}{\rho_{\rm d} \mathfrak{a}_{\rm d}}
    \label{eq_beta}.
\end{equation}
\noindent Here, $\mathcal{C}_{\rm pr} = 1.19 \times 10^{-3}$ kg m$^{-2}$ is the proportionality constant, and $\mathcal{Q}_{\rm pr} \approx 1$ is the scattering coefficient for typical cometary dust \citep{1968ApJ...154..327F,1979Icar...40....1B}. Dust grains subjected to a common value of $\beta$ but released over a range of different epochs form a syndyne, while those released at the same epoch from the cometary nucleus with different $\beta$ form a synchrone.

We computed syndyne-synchrone diagrams for Components A and B of C/2018 F4, which were then compared with our observations (Figure \ref{fig:FP}). The orbital similarity of the two components renders no noticeable differences between their syndyne-synchrone diagrams. However, because of the close separation between the components and their relative positions, the dust feature of Component A around the position angle of the comet's negative heliocentric velocity projected onto the sky plane was strongly concealed by Component B, preventing us from meaningfully determining the dominant dust size of Component A, except for our last observation in 2021 December, when the two components were apparently much further apart from each other but also much fainter than before. We also paid special attention to observations around the time when we were in the orbital plane of the comet and when we were the furthest from the plane. The former condition would collapse all syndynes and synchones into a single line and greatly helps us constrain the dust ejection speed, which is neglected in the Finson-Probstein model. The latter offers us a vantage point where different syndynes and synchrones would be maximally separated from our perspective.

The general dust morphology of Components A and B remained largely unchanged in our observations, despite that the observing geometry, most notably the orbital plane angle, varied significantly (see Figure \ref{fig:vgeo}d). This indicates that the dust environment of the comet was dominated by large-sized particles. Indeed, comparisons between the syndyne-synchrone diagrams and observations consistently yielded that the main tail axis of Component B aligned nicely with syndynes of $\beta \sim 10^{-3}$ (corresponding to dust size $\mathfrak{a}_{\rm d} \sim 1$ mm) having release times no earlier than early 2020. As for Component A, using the last LOOK observation from 2021 December, we found that its dust morphology was also well matched with synchrones having $\beta \sim 10^{-3}$ and release epochs later than 2020 April, the same as what we obtained for Component B from the same observation. Judging from the nearly indistinguishable resemblance in the appearances of the two components around their respective inner regions throughout our observing campaign (see Figures \ref{fig:obs_LOOK} and \ref{fig:obs_UH88}), we infer that the dust environments of the two components were highly alike. Based on the dust release epochs, we deduced that the splitting of the cometary nucleus likely occurred in a time frame between early and mid-2020. Thereafter, both of the nuclear components were ejecting mm-sized dust grains in a protracted manner. The dominance of such large grains and the absence of small grains are qualitatively consistent with the model by \citet{2015A&A...583A..12G} where the small counterparts are retained by interparticle cohesion. Moreover, since the size of dominant dust grains remained unchanged, we prefer that the observed decrease in the effective scattering cross-sections of the two nuclear components (Figure \ref{fig:xs}) was more likely caused by their dwindling activity as the comet receded from the Sun, rather than due to dust grains being removed progressively from the photometric aperture by solar radiation pressure.

We now estimate the dust ejection speed, which was ignored in the syndyne-synchrone model. We noticed that isophotal contours in the inner regions of Components A and B remained largely circular throughout the observed period (see Figure \ref{fig:obs_UH88} as an example), indicative of dominant dust grains being ejected largely isotropically at both nuclear components. The observation from UTC 2020 October 20 was taken closest to the plane-crossing time, granting us an unambiguous side view of the comet. From the observing geometry, we can derive that the out-of-plane width of a dust tail at such an epoch is related to the ejection speed of dust as
\begin{equation}
    V_{\rm ej} = \frac{\tan w_{\perp}}{2 r_{\rm H}}\sqrt{\frac{\beta \mu_{\odot} {\it \Delta} \sin \alpha}{2 \tan \ell}}
    \label{eq_Vper}.
\end{equation}
\noindent Here, $V_{\rm ej}$ is the ejection speed of dust, $w_{\perp}$ is the out-of-plane width of the dust tail at projected angular nucleus distance $\ell$, and $\mu_{\odot} = 3.96 \times 10^{-14}$ au$^3$ s$^{-2}$ is the heliocentric gravitational constant. From the observation, we measured the approximate width of the dust tail to be $w \approx 35\arcsec$ at nucleus distance $\ell \approx 5\arcsec$ for Component B. Substituting, we obtained $V_{\rm ej} \approx 2$ m s$^{-1}$. For Component A, the above method was inapplicable because of the contamination from Component B. Instead, we estimated the dust ejection speed of Component A using the turnaround distance of its dust coma in the sunward direction in the sky plane from
\begin{equation}
    V_{\rm ej} = \frac{\sqrt{2\beta \mu_{\odot} {\it \Delta} \tan \ell \sin \alpha}}{r_{\rm H}}
    \label{eq_Vej},
\end{equation}
\noindent where $\ell$ now refers to the angular nucleus distance of the turnaround point. After inserting the measured $\ell \approx 20\arcsec$ into the above equation, we find $V_{\rm ej} \approx 2$ m s$^{-1}$, which is similar to what we obtained for the dust ejection speed at Component B. Given the measurement uncertainties, these results are possibly no better than order-of-magnitude estimates. Nevertheless, the similarity in the ejection speeds of the dominant dust grains of the two nuclear components makes the two nuclear components appear to be even more homogeneous.

\subsection{Splitting Dynamics}
\label{ss_sd}

\begin{deluxetable}{lccc}
\tablecaption{Best-fitted Orbital Solutions for Components A \& B of C/2018 F4 (PANSTARRS)
\label{tab:orb}}
\tablewidth{0pt}
\tablehead{
\multicolumn{2}{c}{Quantity}  & 
Component A  & 
Component B
}
\startdata
Perihelion distance (au) & $q$
       & 3.4412694(50)
       & 3.4410593(57) \\ 
Eccentricity & $e$
       & 1.0007643(31) 
       & 1.0008969(35) \\ 
Inclination (\degr) & $i$
       & 78.086245(34) 
       & 78.087146(40) \\ 
Argument of perihelion (\degr) & $\omega$
                 & 263.22753(12) 
                 & 263.21357(14) \\ 
Longitude of ascending node (\degr) & ${\it \Omega}$
                 & 26.515932(27) 
                 & 26.515447(32) \\ 
Time of perihelion (TDB)\tablenotemark{$\dagger$} & $t_\mathrm{p}$
                  & 2019 Dec 03.95694(43)
                  & 2019 Dec 03.90181(49) \\
\hline
\multicolumn{2}{l}{Observed arc}
& 2020 Sep 12-2021 Dec 02
& 2020 Sep 12-2021 Dec 02\\
\multicolumn{2}{l}{Number of observations used}
& 199
& 192 \\
\multicolumn{2}{l}{Residual rms (\arcsec)}
& 0.487
& 0.689 \\
\multicolumn{2}{l}{Normalised residual rms}
& 0.739
& 0.861
\enddata
\tablenotetext{\dagger}{The uncertainties are in days.}
\tablecomments{The orbital elements are referenced to the heliocentric J2000 ecliptic, at an osculation epoch of TDB 2021 Dec 2.0 = JD 2459550.5. The reported uncertainties are $1\sigma$ formal errors.}
\end{deluxetable}

In order to explore the splitting event at comet C/2018 F4, we first measured the astrometry of both nuclear components with the Gaia Data Release 2 Catalogue \citep{2018yCat.1345....0G} as the reference for background sources. Although the nonsidereal tracking mode was turned on for our observations of the comet, field stars in the obtained images were not visibly elongated thanks to the apparent motion of the comet, image resolution, and seeing. Thus, we simply treated both nuclear components of the comet and background stars as bidimensional symmetric Gaussians to be fitted. The measurement uncertainties in the astrometric measurements of the nuclear components were properly propagated from errors in the astrometric reduction and centroiding. To extend the observed arcs of the components, we downloaded additional astrometry from the Minor Planet Center Database Search\footnote{\url{https://minorplanetcenter.net/db_search}}.

We then proceeded to compare the orbits of the two nuclear components by means of traditional orbit determination with {\tt FindOrb}\footnote{{The package written by B. Gray is publicly available at \url{https://github.com/Bill-Gray/find_orb}.}}. The astrometry obtained from the Minor Planet Center was debiased and weighted following the methods detailed in \citet{2020Icar..33913596E} and \citet{2017Icar..296..139V}, respectively, whereas ours was simply weighted according to the measured astrometric uncertainties, because of no astrometric uncertainties available for the former. Next, we employed {\tt FindOrb} to fit the orbital elements of each nuclear component, taking into account perturbations from the major planets, Pluto, the Moon, and the 16 most massive asteroids in the main belt with the planetary and lunar ephemeris DE440 \citep{2021AJ....161..105P}, as well as relativistic effects. The observed-minus-calculated ($O-C$) astrometric residuals were largely consistent with the assigned or measured errors. Several measurements whose $O-C$ residuals exceeded the $3\sigma$ level were downweighted accordingly. Nonetheless, this modification did not alter the initial orbital solutions beyond the $1\sigma$ uncertainty levels for both nuclear components. We list and compare the best-fitted orbital solutions in Table \ref{tab:orb}, where the orbital similarity is clearly manifested, strongly indicative of a common origin for the two components.

However, the orbital similarity of the two nuclear components itself is not sufficient to straightforwardly answer the questions of how and when the splitting event occurred. Therefore, we applied the fragmentation model first introduced by \citet{1977Icar...30..574S,1978Icar...33..173S}, which simplifies a splitting event as an instantaneous separation between two nuclear components at some fragmentation epoch, with a component leaving the other one with some relative velocity. The model neglects mutual gravitational interactions between the two components. Despite the similarity in the orbits of the two nuclear components, their orbital elements are statistically different, often indicative of a nonzero relative speed between the components at the time of splitting. 

\begin{figure*}
\begin{center}
\gridline{\fig{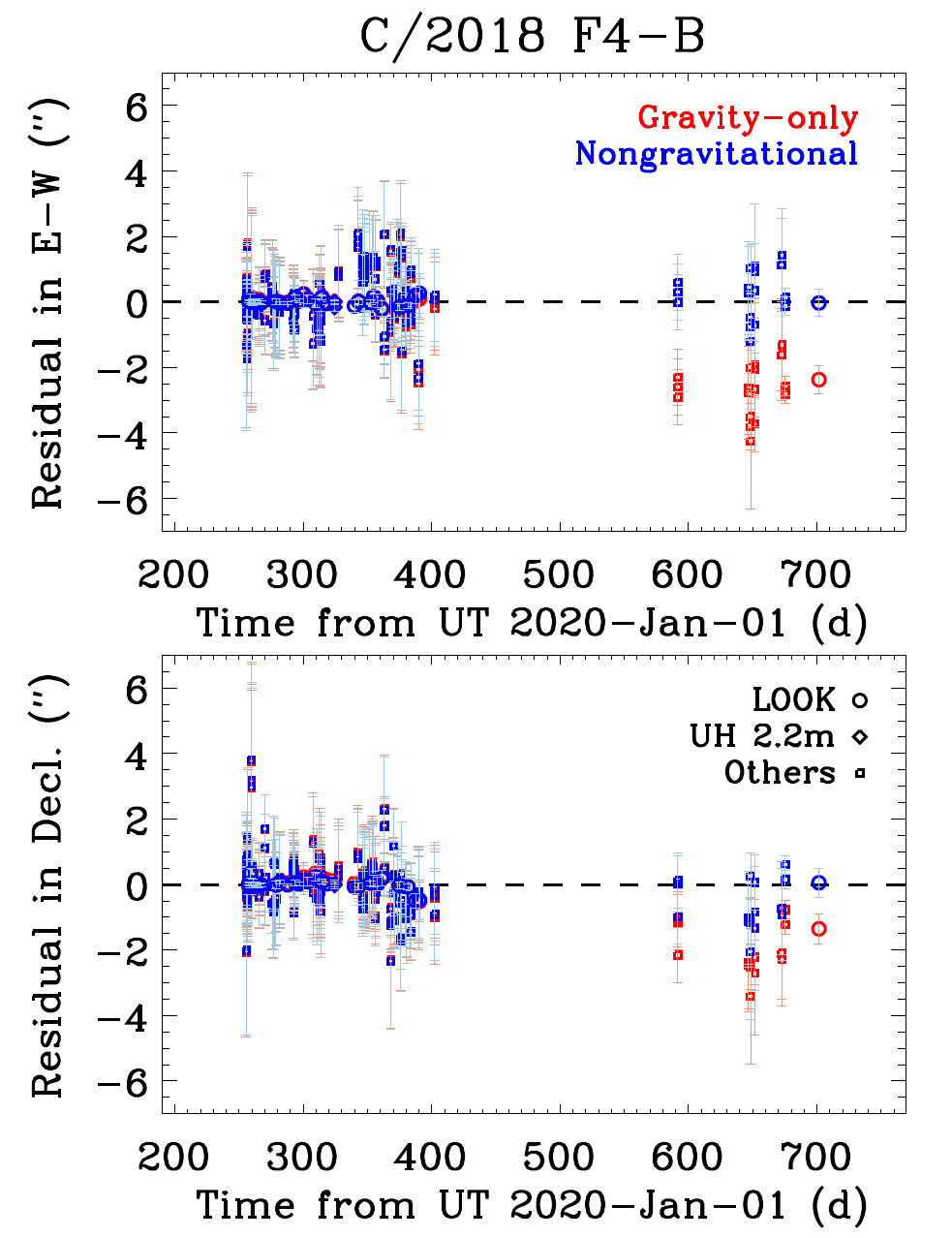}{0.5\textwidth}{(a)}
\fig{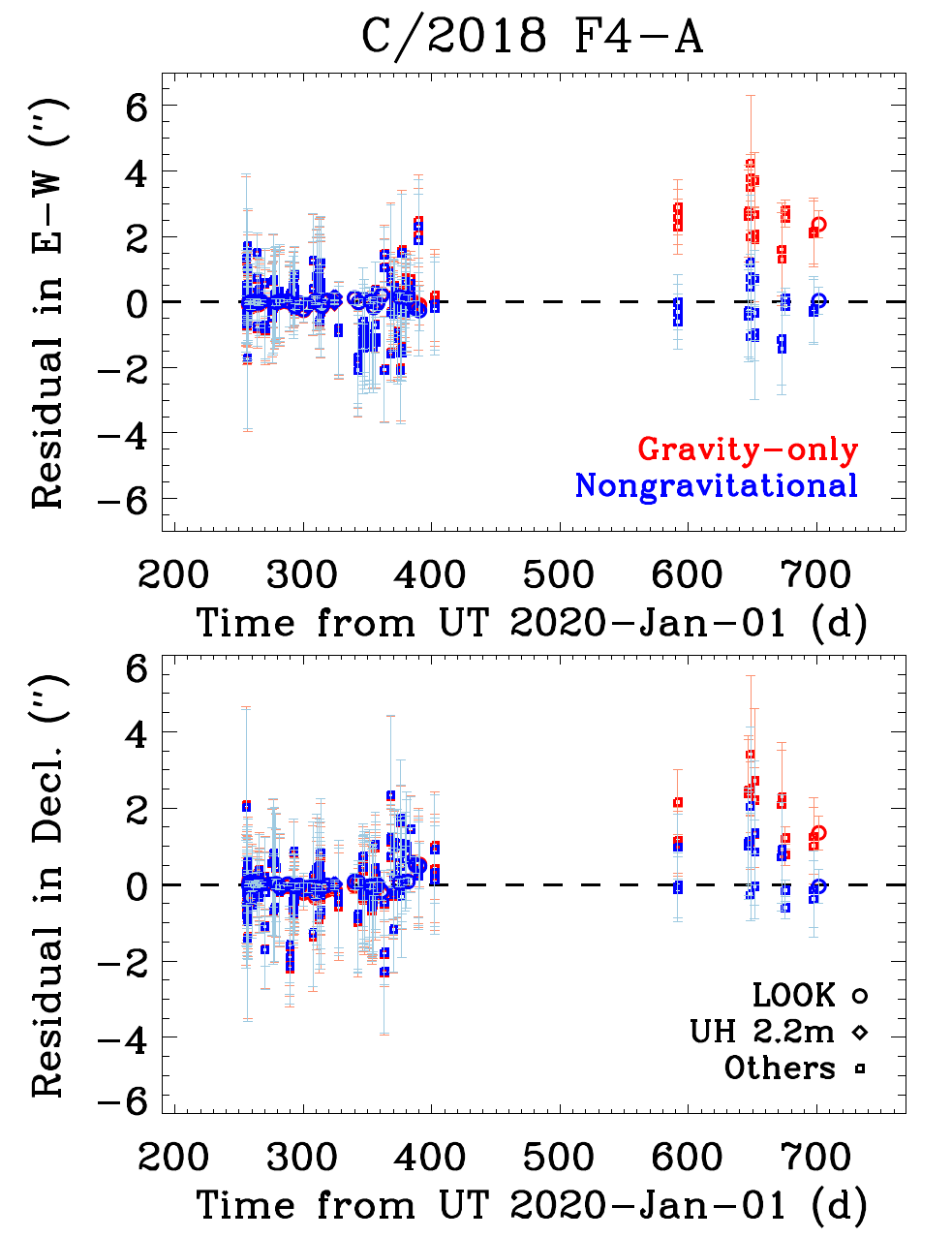}{0.5\textwidth}{(b)}}
\caption{
Astrometric O-C residuals in the gravity-only model (red) and the nongravitational one (blue) for the splitting at C/2018 F4. The left two panels (a) are residuals in the J2000 equatorial east-west (top) and decl. (bottom) directions for the scenario where Component A was treated as the primary and astrometry of Component B was fitted, whereas the right two panels (b) are for the scenario where Component B was assumed to be the primary and we instead fitted the observations of Component A. In each panel, data points in different symbols correspond to observations from different telescopes, but those not from our measurements are not further discriminated for clarity and brevity. It is evident that the nongravitational model offers a more satisfactory solution than the gravity-only model in both scenarios, as the systematic trend primarily in the post-conjunction astrometry is only present in the gravity-only model.
\label{fig:res}
} 
\end{center} 
\end{figure*}

We employed our code, which incorporates the Levenberg–Marquardt optimisation code {\tt MPFIT} \citep{2009ASPC..411..251M} and the same gravitational perturbers with DE440 alongside relativistic corrections were included, to study the splitting at comet C/2018 F4. Previously we have exploited the code to delve into the fragmentation events at active asteroids P/2016 J1 (PANSTARRS) and 331P/Gibbs \citep{2017AJ....153..141H,2022AJ....164..236H}. We first regarded C/2018 F4-A as the primary component of the comet and fitted the relative astrometry of C/2018 F4-B with respect to the former. In cases of no simultaneous measurements and only the astrometry of Component B is available, the relative astrometry was computed together with the obtained orbital elements of C/2018 F4-A listed in Table \ref{tab:orb}. Initially, we attempted to find a best-fit solution to the fragmentation epoch $t_{\rm frg}$ and separation velocity decomposed into radial, transverse, and normal (RTN) components referenced to the primary that would minimise the goodness of fit for the whole observed arc of Component B spanning from 2020 September 12 to 2021 December 2. However, regardless of our initial guessed values, the solution (here termed Solution I of the gravity-only model) always converged to a statistically similar result containing an unacceptably strong systematic trend in the $O-C$ astrometric residuals for observations starting from 2021 August when the comet became observable again after the solar conjunction in 2021 (see Figure \ref{fig:res}a). We then fitted the observations before and after the solar conjunction separately, thereby obtaining much better solutions (termed Solutions II and III, respectively, of the gravity-only model) without any conspicuous systematic trend in the $O-C$ astrometric residuals, characterised by their goodness-of-fit values both being clearly smaller than that of Solution I (see Table \ref{tab_split1}). While separating the astrometry into two distinct arcs helped circumvent the systematic trend in the astrometric residuals, we are aware that the normal components of the best-fitted separation velocities in Solutions II and III do not agree statistically with each other, suggestive of problems in the gravity-only fragmentation model. 

Thus, we decided to further take into account nongravitational effects, which may arise from anisotropic mass-loss activity. Because the fragmentation model in essence requires relative positions between the secondary and the primary components, technically the nongravitational acceleration yielded by the model is referenced with respect to the primary. Given the orbit of the comet (perihelion distance $q = 3.4$ au), we do not posit that nongravitational effects at C/2018 F4 would be driven by sublimation of water ice, because of the low equilibrium temperature. Therefore, the nongravitational force model by \citet{1973AJ.....78..211M} based on isothermal water-ice sublimation is most likely inapplicable. Rather, we adopted a nongravitational force model varying with $r_{\rm H}^{-2}$, following \citet{1977Icar...30..574S,1978Icar...33..173S}, who incorporated nongravitational effects in the radial direction only. However, we found through testing that adding a radial differential nongravitational acceleration to be fitted was insufficient to completely remove the strong systematic trend in the $O-C$ astrometric residuals for the full observed arc, although the goodness of fit was improved. Thus, we further incorporated the transverse and normal components of the differential nongravitational acceleration as additional free parameters in the fragmentation model, whereby the systematic trend was eliminated successfully, with the $O-C$ residuals all well consistent with the measurement uncertainties (see Figure \ref{fig:res}a). We tabulate the best-fit parameters of the nongravitational fragmentation model in Table \ref{tab_split1}.

\begin{deluxetable*}{l@{\extracolsep{\fill}}C@{\extracolsep{\fill}}cccc}
\tabletypesize{\footnotesize}
\tablecaption{Best-fit Fragmentation Models of C/2018 F4 (PANSTARRS): Separation of Component B from Component A
\label{tab_split1}}
\tablewidth{0pt}
\tablehead{
\multicolumn{2}{c}{Quantity}  & 
\multicolumn{3}{c}{Gravity-only Model} &
\multicolumn{1}{c}{Nongravitational Model} \\
\cmidrule(lr){3-5} \cmidrule(lr){6-6}
&&
\colhead{Solution I} &
\colhead{Solution II} &
\colhead{Solution III} &
\colhead{$\sim r_{\rm H}^{-2}$}
}
\startdata
Fragmentation epoch (TDB)\tablenotemark{$\dagger$} & $t_{\rm frg}$
       & 2020 May $2.8 \pm 1.3$
       & 2020 Apr $26.8 \pm 1.7$
       & 2020 Jan $23 \pm 84$
       & 2020 Apr $22.3 \pm 8.3$ \\ 
\multicolumn{2}{l}{Separation velocity (m s$^{-1}$)} &&&\\
& $\Delta V_{\rm R}$
       & $+3.145 \pm 0.015$
       & $+3.031 \pm 0.022$
       & $+2.45 \pm 0.75$
       & $+2.93 \pm 0.16$ \\
& $\Delta V_{\rm T}$
       & $-0.577 \pm 0.023$
       & $-0.500 \pm 0.025$
       & $+0.41 \pm 0.62$
       & $-0.52 \pm 0.17$ \\ 
& $\Delta V_{\rm N}$
       & $-0.1526 \pm 0.0023$
       & $-0.1562 \pm 0.0024$
       & $+0.143 \pm 0.015$
       & $-0.363 \pm 0.011$ \\ 
\multicolumn{2}{l}{Differential NG parameter (au d$^{-2}$)\tablenotemark{$\ddagger$}} &&&\\
& $\Delta A_{\rm R}$ & 0 & 0 & 0
      & $\left(+2.69 \pm 0.37\right) \times 10^{-8}$ \\
& $\Delta A_{\rm T}$ & 0 & 0 & 0
      & $\left(+1.17 \pm 0.73\right) \times 10^{-8}$ \\
& $\Delta A_{\rm N}$ & 0 & 0 & 0
      & $\left(+2.04 \pm 0.13\right) \times 10^{-8}$ \\
\hline
\multicolumn{2}{l}{Observed arc}
& 2020 Sep 12-2021 Dec 02
& 2020 Sep 12-2021 Feb 06
& 2021 Aug 14-2021 Dec 02
& 2020 Sep 12-2021 Dec 02 \\
\multicolumn{2}{l}{Number of observations used}
& 192
& 173
& 19
& 192 \\
\multicolumn{2}{l}{Normalized residual rms}
& 1.364
& 0.804
& 0.755
& 0.710
\enddata
\tablenotetext{\dagger}{The corresponding uncertainties are in days.}
\tablenotetext{\ddagger}{Differential nongravitational parameter.}
\tablecomments{The reported uncertainties are $1\sigma$ formal errors. Here, we treat Component A as the primary nucleus and fitted the astrometric observations of Component B. In each solution, the obtained separation velocity and differential nongravitational parameter of Component B decomposed into the RTN directions are referenced with respect to Component A at the corresponding fragmentation epoch. In the gravity-only model, the differential nongravitational parameter is held fixed at zero.}
\end{deluxetable*}

\begin{deluxetable*}{l@{\extracolsep{\fill}}C@{\extracolsep{\fill}}cccc}
\tabletypesize{\footnotesize}
\tablecaption{Best-fit Fragmentation Models of C/2018 F4 (PANSTARRS): Separation of Component A from Component B
\label{tab_split2}}
\tablewidth{0pt}
\tablehead{
\multicolumn{2}{c}{Quantity}  & 
\multicolumn{3}{c}{Gravity-only Model} &
\multicolumn{1}{c}{Nongravitational Model} \\
\cmidrule(lr){3-5} \cmidrule(lr){6-6}
&&
\colhead{Solution I} &
\colhead{Solution II} &
\colhead{Solution III} &
\colhead{$\sim r_{\rm H}^{-2}$}
}
\startdata
Fragmentation epoch (TDB) & $t_{\rm frg}$
       & 2020 May $2.8 \pm 1.3$
       & 2020 Apr $26.6 \pm 1.6$
       & 2020 Jan $23 \pm 82$
       & 2020 Apr $22.5 \pm 8.1$ \\ 
\multicolumn{2}{l}{Separation velocity (m s$^{-1}$)} &&&\\
& $\Delta V_{\rm R}$
       & $-3.146 \pm 0.015$
       & $-3.028 \pm 0.022$
       & $-2.45 \pm 0.74$
       & $-2.94 \pm 0.15$ \\
& $\Delta V_{\rm T}$
       & $+0.575 \pm 0.022$
       & $+0.497 \pm 0.025$
       & $-0.43 \pm 0.60$
       & $+0.53 \pm 0.16$ \\ 
& $\Delta V_{\rm N}$
       & $+0.1522 \pm 0.0023$
       & $+0.1558 \pm 0.0023$
       & $-0.143 \pm 0.015$
       & $+0.363 \pm 0.011$ \\ 
\multicolumn{2}{l}{Differential NG parameter (au d$^{-2}$)} &&&\\
& $\Delta A_{\rm R}$ & 0 & 0 & 0
      & $\left(-2.72 \pm 0.36\right) \times 10^{-8}$ \\
& $\Delta A_{\rm T}$ & 0 & 0 & 0
      & $\left(-1.20 \pm 0.72\right) \times 10^{-8}$ \\
& $\Delta A_{\rm N}$ & 0 & 0 & 0
      & $\left(-2.04 \pm 0.13\right) \times 10^{-8}$ \\
\hline
\multicolumn{2}{l}{Observed arc}
& 2020 Sep 12-2021 Dec 02
& 2020 Sep 12-2021 Feb 06
& 2021 Aug 14-2021 Dec 02
& 2020 Sep 12-2021 Dec 02 \\
\multicolumn{2}{l}{Number of observations used}
& 199
& 178
& 21
& 199 \\
\multicolumn{2}{l}{Normalized residual rms}
& 1.374
& 0.834
& 0.722
& 0.734
\enddata
\tablecomments{Same as Table \ref{tab_split1}, except that Component B is now treated instead as the primary and we fitted the astrometry of Component A. In each solution, the separation velocity and the differential nongravitational parameter of Component A are expressed in terms of the RTN components relative to Component B at the instant of splitting.}
\end{deluxetable*}

To demonstrate the reliability and robustness of the results, we repeated our aforementioned steps but instead treated C/2018 F4-B as the primary and fitted the relative astrometry of C/2018 F4-A. We again had to fit the observations separately with the gravity-only model, otherwise the resulting $O-C$ astrometric residuals in the post-conjunction observations would contain a strong systematic trend well beyond the noise level (see Figure \ref{fig:res}b). Thus, the nongravitational model was invoked, with which we obtained almost identical best-fit results, except that the signs of RTN components of the separation velocity and differential nongravitational parameter are flipped, because here the roles between the two nuclear components are swapped (see Table \ref{tab_split2}).

In addition, we were concerned about how tailward biases in our astrometric measurements of the two components of comet C/2018 F4, if present, would influence the best-fit results. Such biases have been previously observed to exist for astrometric measurements of comets whose comae are asymmetric due to solar radiation pressure and/or anisotropic cometary activity, and therefore extrapolating to zero-aperture astrometry would be desirable \citep[e.g.,][]{2020AJ....160...91H}. However, because neither of the nuclear components of C/2018 F4 had sufficiently high signal-to-noise ratios, nor appeared strongly elongated (see Figures \ref{fig:obs_LOOK} and \ref{fig:obs_UH88}), we found that the zero-aperture astrometry had enormous uncertainties often exceeding $\sim\!1\arcsec$. Therefore, we saw no benefit in using zero-aperture astrometry. Rather, we utilised astrometry obtained with different aperture sizes, only to find that the nongravitational model would be needed to fully eliminate the systematic trend in the $O-C$ astrometric residuals, and that the obtained best-fit parameters are all consistent within the noise level. Now we can firmly conclude that comet C/2018 F4 split into two major components most likely in late April 2020, broadly consistent with the result from our syndyne-synchrone computation (Section \ref{ss_morph}), at a heliocentric distance of $\sim\!3.7$ au with a separation speed of $3.00 \pm 0.18$ m s$^{-1}$ between the two components, mostly in the radial direction. Furthermore, the best-fit models reveal that Component B was subject to a nongravitational acceleration stronger than the one of Component A, statistically important in a plane perpendicular to the orbital transverse direction. Here, we feel the necessity to comment on the detection of the differential nongravitational effect between the two nuclear components of the comet.

At first glance, the detection of the differential nongravitational acceleration may seem to contradict our results from traditional orbit determination where gravity-only solutions were found to be sufficient, but in actuality, it does not. The reason is that the common origin of Components A and B is completely neglected in the traditional orbit determination, and yet this turns out to be the most important constraint of all. If we also incorporate the nongravitational model in {\tt FindOrb}, there will be no statistically confident detection of the nongravitational parameters except in the normal direction for C/2018 F4-B, $A_{\rm N} = \left(+2.01 \pm 0.50\right) \times 10^{-8}$ au d$^{-2}$. However, the resulting differential nongravitational parameter in the normal direction is still statistically insignificant, as the uncertainty of the other component is dominant. Nonetheless, the result is still in agreement with the one from the nongravitational fragmentation model within the $1\sigma$ level. We thus believe that the differential nongravitational acceleration between the two components of C/2018 F4 is authentic. This find possibly indicates a smaller size of Component B compared to Component A of C/2018 F4, but it is also possible that this was a consequence of the mass loss at Component B being more anisotropic.  

\section{Discussion}
\label{sec_disc}


Here we first discuss the nucleus size of comet C/2018 F4 using two different methods. First, we constrain the size of each component from the cross-section estimates in Section \ref{ss_phot}. For each component, the upper limit is given by the radius of the equal-area circle using the earlier observations, whereas the lower limit can be obtained from the last observation, assuming the effective scattering cross-section is wholly due to the ejected dust. Thus, we obtain a common upper limit to the nucleus radii of both components to be $R_{\rm n} \la \sqrt{{\it \Xi}_{\rm e} / \pi} \approx 14 \pm 2$ km. As for the lower limits, we find $R_{\rm n} \ga \left[{\it \Xi}_{\rm e} \rho_{\rm d} \bar{\mathfrak{a}}_{\rm d} / \left(2 \pi \rho_{\rm n} \right) \right]^{1/3} \sim 60$ m for both components, where $\rho_{\rm n} = 0.5$ g cm$^{-3}$ is adopted as the nominal bulk density of cometary nuclei \citep[][and citations therein]{2019SSRv..215...29G}. Given that cometary nuclei were reported to have surface geometric albedo values falling within a narrow range of 0.02-0.06 \citep{{2004come.book..223L}}, we do not posit that our constraints assuming $p_r = 0.05$ can deviate from the reality by over $\sim$60\%.

The derived dust ejection speed in Section \ref{ss_morph} and the separation speed between Components A and B in Section \ref{ss_sd} can be also useful to place an upper limit to the nucleus size for each of the nuclear components, as we postulate that the escape speed at each component should be no greater than these speeds. Setting the escape speed equal to the smaller ejection speed of the dominant dust grains, we find that the upper limit to the nucleus radius of each component is $R_{\rm n} \la V_{\rm ej} / \sqrt{8\pi G \rho_{\rm n} / 3} \approx 4$ km, where $G = 6.67 \times 10^{-11}$ m$^3$ kg$^{-1}$ s$^{-2}$ is the gravitational constant. Therefore, combined with the estimates from the first method and taking various uncertainties into account, the volume-equivalent diameter of the pre-split nucleus of the comet should lie within a range of $\sim\!0.1$-10 km.

The similarities in the estimated nuclear sizes and the colours of the two nuclear components prompt us to speculate that the pre-split nucleus of C/2018 F4 might have been bilobate in shape, resembling, for instance, 67P's nucleus, which was found to consist of two lobes in colours highly similar to each other jointed by a narrow neck \citep{2015Sci...347a1044S}, and that the nucleus for some reason broke its neck region and liberated the two lobes as the two nuclear components we observed. In fact, the bilobate shape appears to be common for cometary nuclei, judging from the fact that five out of seven in situ observed cometary nuclei were found to be of such shape \citep{2016Natur.534..352H}, possibly formed from mergers of two distinct objects \citep{2015Sci...348.1355J,2017A&A...597A..62J,2018NatAs...2..379S} or sublimative torques \citep{2021PSJ.....2...14S}. 

\begin{figure*}
\begin{center}
\gridline{\fig{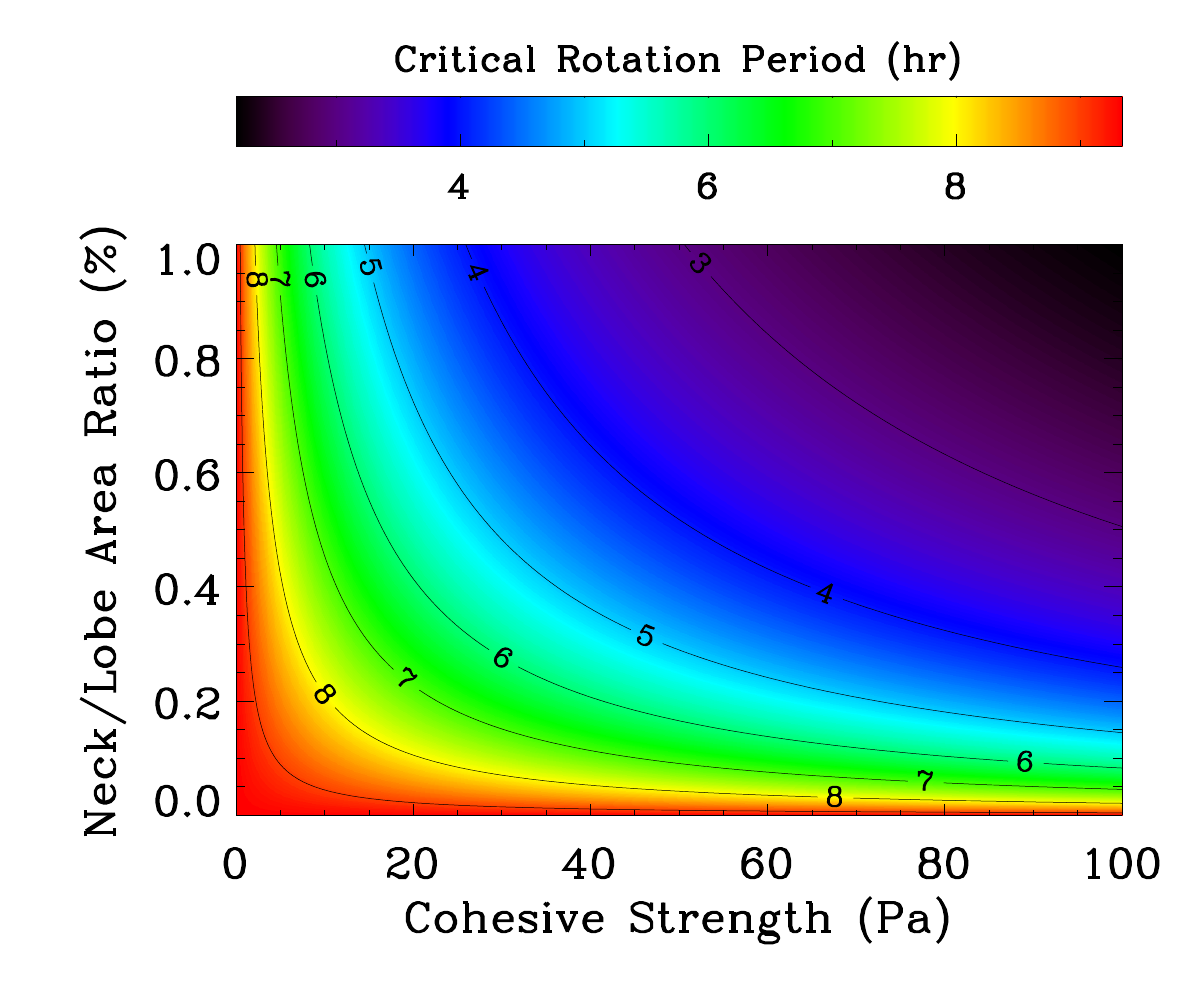}{0.5\textwidth}{(a)}
\fig{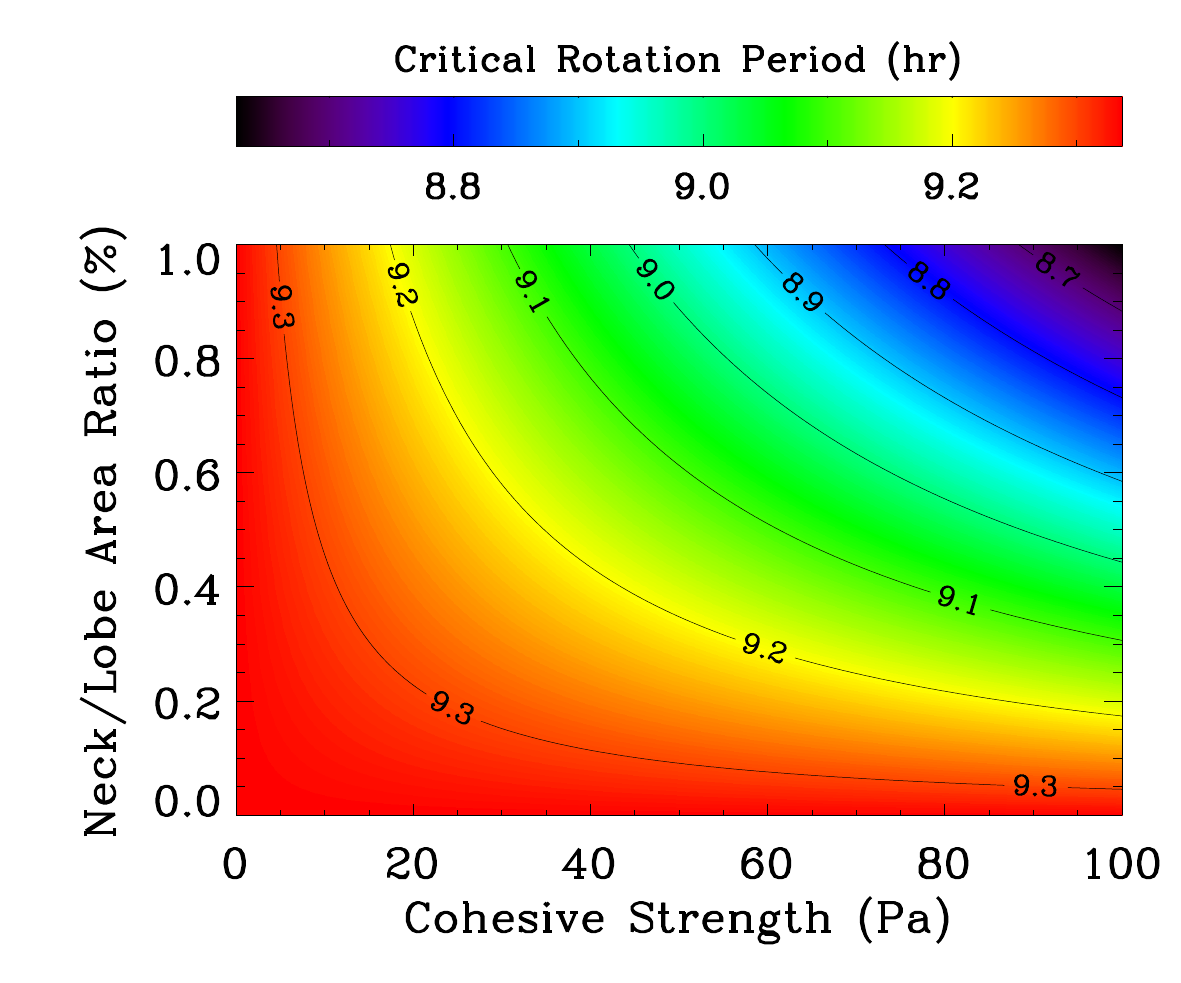}{0.5\textwidth}{(b)}}
\caption{
Critical rotation period versus cohesive strength and neck-to-lobe area ratio. We assume lobe radii of 100 m and 1 km for the left and right panels, respectively. Given some nucleus size, the general trend is that the larger the cohesive strength and the neck-to-lobe area ratio, the shorter the critical rotation period. For a larger nucleus, the influence of cohesion is less important.}
\label{fig:Prot}
\end{center} 
\end{figure*}

As for the mechanism by which the two lobes separated, we first suggest the possibility of rotational breakup. To see why, we first assume the pre-split nucleus to be a twin of identical solid spheres of uniform mass density resting on each other. Separation between the two components occurs when the rotation period of the nucleus is so short that the centrifugal force starts to rival their mutual gravitational pull. As such, we can derive rotation period $P_{\rm rot} \le 2\sqrt{3\pi / \left(\rho_{\rm n} G\right)} \approx 9$ hr to satisfy the condition of separation. If cohesion at the region where the two lobes contact is included, even shorter rotation periods will be needed for the split. In this scenario, we approximate the original nucleus as two identical spherical caps merged at their bases forming the neck region, within the plane of which lies the spin axis of the nucleus. Breakup of the two lobes occurs when the difference between the centrifugal force and the mutual gravity pull exceeds the cohesion at the neck. Let us set the area of the neck to be much smaller than the overall surface area of each lobe. To split such a nucleus, the rotation period should satisfy
\begin{equation}
P_{\rm rot} \la 2\pi R_{\rm n}\sqrt{\frac{3 \rho_{\rm n}}{9 \xi \sigma + \pi G \rho_{\rm n}^2 R_{\rm n}^2}}
\label{eq_Prot},
\end{equation}
\noindent where $\xi \ll 1$ is the neck-to-lobe area ratio, and $\sigma$ is the cohesive strength at the neck. We show the critical rotation period as a function of the aforementioned two variables in Figure \ref{fig:Prot}, with a lobe radius of 100 m for the left panel, and 1 km for the right one. Here, we only consider values of global cohesive strength within the range typical for cometary nuclei \citep[$\la\!100$ Pa;][]{2019SSRv..215...29G}. As we can see, given some fixed nucleus size, a shorter critical rotation period is needed for the nucleus having greater cohesive strength at the neck region and a broader neck to split, which is not counterintuitive. We also find that the effect from a nonzero cohesion at the neck becomes less important for greater nucleus sizes. Although the rotation period of the pre-split cometary nucleus of C/2018 F4 is unknown, we still posit that the nucleus could split due to rotational instability, judging from the fact that there are nuclei of long-period comets with rotation periods falling within the obtained regime \citep[and citations therein]{{2023arXiv230409309K}}. 

Alternatively, C/2018 F4 might have been a binary comet resembling main-belt comet 288P \citep{2017Natur.549..357A}, in which case Components A and B might have resulted from the separation of the binary system due to some dynamical instability. Assuming the spin statuses of both nuclear components remain unchanged before and after the splitting, in the case of contact binary, we derive that an increase in the total free specific energy of the binary system greater than the total orbital energy, namely,
\begin{equation}
\Delta \mathscr{E} \ge \frac{1}{6}\pi G \rho_{\rm n} R_{\rm n}^2
\label{eq_DE},
\end{equation}
\noindent may cause a mutual escape of the system. Substituting, we obtain $\Delta \mathscr{E} \ga 10^{-4}$-$10^{-1}$ J kg$^{-1}$. The larger the nucleus, the greater the specific energy increase will be needed. The minimum specific energy increase can be estimated from the scenario where the separation distance between the two nuclear components is the Hill radius, thereby giving
\begin{equation}
\min \left\{\Delta \mathscr{E}\right\} = \frac{\pi G \rho_{\rm n} R_{\rm n}^2}{3q} \left(\frac{9M_{\odot}}{4\pi\rho_{\rm n}}\right)^{1/3}
\label{eq_DE_min}.
\end{equation}
\noindent Here, $M_{\odot} \approx 2 \times 10^{30}$ kg is the mass of the Sun, and $q = 3.4$ au is the perihelion distance of the comet. Substitution with the obtained nucleus size range yields a minimum specific energy increase of $\sim\!10^{-6}$-$10^{-3}$ J kg$^{-1}$, to cause a mutual escape of the binary system. On the other hand, our fragmentation model (see Section \ref{ss_sd}) is informative of the specific energy change, which can be calculated from
\begin{equation}
\Delta \mathscr{E} \approx {\bf V} \cdot \left(\Delta {\bf V}_{\rm R} + \Delta {\bf V}_{\rm T} + \Delta {\bf V}_{\rm N}\right)
\label{DE_sd}
\end{equation}
\noindent as the change in the specific kinetic energy of the two nuclear components of C/2018 F4. Here, ${\bf V}$ is the heliocentric velocity of C/2018 F4 instantaneously before the splitting, and the RTN separation velocities are all vectors. The result is $\Delta \mathscr{E} = \left(5.3 \pm 2.8 \right) \times 10^3$ J kg$^{-1}$, which unsurprisingly falls within the aforementioned ranges, thus making the breakup of a binary system possible. Furthermore, it is noteworthy that the obtained value from the best-fit fragmentation model does not outnumber those reported for cometary outbursts, such as 15P/Finlay, 17P/Holmes, and 332P/Ikeya-Murakami \citep[$\sim\!10^4$-$10^{5}$ J kg$^{-1}$;][and citations therein]{2016AJ....152..169I}. Thus, it seems reasonable that bursting cometary activity at C/2018 F4 might have caused the separation between its two nuclear components.

Unfortunately, because we have no observations whatsoever prior to the discovery of the splitting of the comet, by no means can we judge which was the more plausible mechanism whereby C/2018 F4 split. The fragmentation model alone does not uniquely imply a specific mechanism of the splitting, as neither the separation speed nor the differential nongravitational acceleration obtained from the best fit is more than mediocre if compared to previous results for other split comets \citep[and citations therein]{2004come.book..301B}. Moreover, C/2018 F4 split in a fashion not much different from other split comets of Type A categorized in \citet{2004come.book..301B}, where the comet splits into two (or only a few more) components, whereas a split comet of Type B disintegrates into many ($\ga\!10$) pieces and no primary component can be identified. Had there been data of pre-split C/2018 F4, there would be a reasonable chance to allow for measuring the rotation of its nucleus and monitoring the temporal evolution of the cometary activity, the results of which would be diagnostic of the splitting mechanism. Nevertheless, we formulate that future remote observations of split comets with two components of similar properties may provide a feasible manner for us to constrain the binarity or bilobate shape fraction of cometary nuclei, which will help us better understand the formation processes of planetesimals in the early solar system.


\section{Summary}
\label{sec_sum}

We obtained observations of long-period comet C/2018 F4 (PANSTARRS) immediately after the splitting event was detected, from which two co-moving nuclear components were seen. Our study obtained the following key findings.

\begin{enumerate}
    \item The two nuclear components remained highly similar to each other in terms of brightness, colour, and dust morphology over the course of our observing campaign. The colour of the comet, $g-r=0.52 \pm 0.04$ and $r-i = 0.17 \pm 0.06$ for Component A, and $g-r=0.54 \pm 0.04$ and $r-i=0.08 \pm 0.06$ for Component B, is consistent with those of many other long-period comets.
    \item From our photometry, we estimated their individual total effective scattering cross-sections, which in general declined with time, more likely resulting from diminishing cometary activity as the comet receded from the Sun. Assuming geometric albedo $0.05$ and bulk mass density $\sim\!1$ g cm$^{-3}$ for the dominant dust grains, the average mass-loss rates were $\sim\!9 \pm 4$ kg s$^{-1}$ for both nuclear components.
    \item The syndyne-syncrhone computation suggested that the dust morphology of both nuclear components was dominated by mm-sized dust grains ejected in a protracted manner no earlier than early 2020. Using the observation around the plane-crossing time, we estimated the dust ejection speed to be $\sim\!2$ m s$^{-1}$.
    \item We obtained a similar nucleus radius for the two nuclear components, in a range from $\sim\!60$ m to 4 km, based on analyses of photometry, dust morphology, and split dynamics. Thus, the volume-equivalent diameter of the pre-split nucleus was estimated to be between $\sim\!0.1$ km and $10$ km.
    \item Our fragmentation model yielded that the two nuclear components split at a relative speed of $3.00 \pm 0.18$ m s$^{-1}$ largely in the radial direction in late April 2020, implying a specific energy change of $\left(5.3 \pm 2.8 \right) \times 10^3$ J kg$^{-1}$. After the separation, Component B was subjected to a stronger nongravitational acceleration than Component B, with the difference statistically important in both the radial and normal directions.
    \item We postulate that the pre-split C/2018 F4 might either have had a bilobate nucleus resembling the one of 67P for instance, or have been a binary system like main-belt comet 288P. Possible splitting mechanisms include rotational breakup and bursting cometary activity. We suggest remote observations of split comets be a feasible manner to constrain the binarity or bilobate shape fraction of cometary nuclei.

\end{enumerate}

\begin{acknowledgements}
We thank the anonymous reviewer for insightful comments on our manuscript, Bill Gray for implementing {\tt FindOrb}, David Tholen for sharing his codes for astrometry, and Gianpaolo Pizzetti for making publicly available {\tt AstroMagic}, which was employed in combination with the aforementioned codes for astrometry. MTH is thankful to UCLA EPSS Department for allowing the temporary use of their IDL license as an alumnus and to the Science and Technology Development Fund, Macau SAR, for the support through grant No. 0016/2022/A1. MSPK is supported by NASA grant 80NSSC21K0164.
This work makes use of observations from the Las Cumbres Observatory global telescope network.  Observations with the LCOGT 1m telescopes were obtained as part of the LCO Outbursting Objects Key (LOOK) Project (KEY2020B-009). The authors wish to recognize and acknowledge the very significant cultural role and reverence that the summit of Mauna Kea has always had within the indigenous Hawaiian community.  We are most fortunate to have the opportunity to conduct observations from this mountain. This research has made use of data and services provided by the International Astronomical Union's Minor Planet Center, JPL Horizons, and NASA's Astrophysics Data System Bibliographic Services.

\noindent Data Access: Raw Data Products (and Basic Calibrated Data products as described in Section~\ref{sec_obs}) supporting this study are available from the LCO Science Archive at \url{https://archive.lco.global} using the KEY2020B-009 proposal code following a 12 month embargo/proprietary period.  This work also made use of MAST CasJobs and its hosted ATLAS Refcat2 database \dataset[https://doi.org/10.17909/t9-2p3r-7651]{https://doi.org/10.17909/t9-2p3r-7651}.
\end{acknowledgements}
\vspace{5mm}
\facilities{LCOGT, UH 2.2m}

\software{{\tt AstroMagic}, {\tt calviacat} \citep{2019zndo...2635841K}}, {\tt FindOrb}, {\tt IDL}, {\tt MPFIT} \citep{2009ASPC..411..251M}




\appendix

\section{Correction for Aperture Photometry of A Pair of Partially Overlapping Comae}
\label{sec_apndx}

Here we detail our derivation of the correction for aperture photometry centred on an azimuthally symmetric coma (1), in part contaminated by another azimuthally symmetric coma (2) whose centre is separated from the centroid by an angular distance of $s$. For simplicity, we regard both comae to be in steady state without any influence from solar radiation pressure, and therefore, both have surface brightness profiles inversely proportional to the distance from their respective optocentres.

The measured total flux of coma 1 ($F'_1$) within a photometric aperture of $\rho$ in radius centred at its optocentre is then comprised of two parts, contribution from coma 1 ($F_1$, the true brightness we wish to recover) and that from coma 2 (the contamination we wish to get rid of):
\begin{equation}
F'_{1} = 2 \int\limits_{0}^{\pi} \int\limits_{0}^{\rho} \left(\frac{k_1}{r} + \frac{k_2}{r'} \right) r~{\rm d}r~{\rm d}\theta
\label{eq_Fobs},
\end{equation}
\noindent where $k_j$ ($j=1,2$) is the proportionality constant for the surface brightness profile of the $j$-th coma, $r$ and $r'$ are respectively the distances from the optocentres of comae 1 and 2, and $\theta$ is the azimuthal angle with respect to the position angle of the straight line connecting optocentres of the two comae.

From the cosine law, we can write
\begin{equation}
r' = \sqrt{r^2 - 2 r s \cos \theta + s^2}
\label{eq_coslaw}.
\end{equation}
\noindent Substituting, and after some algebra, Equation (\ref{eq_Fobs}) then becomes
\begin{equation}
F'_{1} = 2\pi k_1 \rho + 2k_2 s \int\limits_{0}^{\pi} \int\limits_{0}^{\rho/s} \frac{u~{\rm d}u~{\rm d}\theta}{\sqrt{u^2 - 2u \cos \theta + 1}}
\label{eq_Fobs2},
\end{equation}
\noindent in which we set $u \equiv r/s$. In our aperture photometry, we limited our aperture sizes such that the optocentre of coma 2 always remains exterior to the aperture, corresponding to $u \in (0,1)$. We denote the integral by $\mathcal{J}$, and proceed to evaluate its value by integrating over $u$ first and then over $\theta$. The result is
\begin{equation}
\mathcal{J} = \left(1 - \frac{\rho}{s} \right) E\left(-\frac{4 \rho s}{\left(s - \rho\right)^2} \right)
- \left(1 + \frac{\rho}{s} \right) K\left(-\frac{4 \rho s}{\left(s - \rho\right)^2} \right)
\label{eq_J}.
\end{equation}
\noindent Here, $K$ and $E$ are the complete elliptic integrals of the first and second kinds, respectively. Then the ratio between the measured (contaminated) and actual (corrected) fluxes of coma 1 is
\begin{equation}
\frac{F'_1}{F_1} = 1 + \frac{\mathcal{J} k_2}{\pi k_1} \left(\frac{s}{\rho} \right)
\label{eq_Fcor},
\end{equation}
\noindent the reciprocal of which is referred to as the flux correction factor for coma 1. By interchanging the subscripts indices in Equation (\ref{eq_Fobs2}), we can also obtain the expression for the measured total flux of coma 2. Thus, we can solve for the two proportionality constants to be
\begin{equation}
k_1 = \frac{\pi \rho F'_1 - \mathcal{J}s F'_2}{2\left(\pi^2 \rho^2 - \mathcal{J}^2 s^2 \right)}
\label{eq_k1},
\end{equation}
\noindent and
\begin{equation}
k_2 = \frac{\pi \rho F'_2 - \mathcal{J} s F'_1}{2\left(\pi^2 \rho^2 - \mathcal{J}^2 s^2 \right)}
\label{eq_k2}.
\end{equation}
\noindent Thereby, the uncontaminated flux of coma 1 can be immediately restored:
\begin{equation}
F_1 = \pi \rho \left(\frac{\pi \rho F'_1 - \mathcal{J} s F'_2}{\pi^2 \rho^2 - \mathcal{J}^2 s^2}\right)
\label{eq_Freal},
\end{equation}
\noindent and the one of coma 2 can be also conveniently obtained by swapping the subscript indices. The flux correction factor can then be derived as
\begin{equation}
\frac{F_1}{F'_1} = \frac{\pi^2 \rho^2}{\pi^2 \rho^2 - \mathcal{J}^2 s^2} \left[1 - \left(\frac{\mathcal{J}s}{\pi \rho}\right) \frac{F'_2}{F'_1} \right]
\label{eq_Fcor2}.
\end{equation}

\begin{figure}
\epsscale{.7}
\begin{center}
\plotone{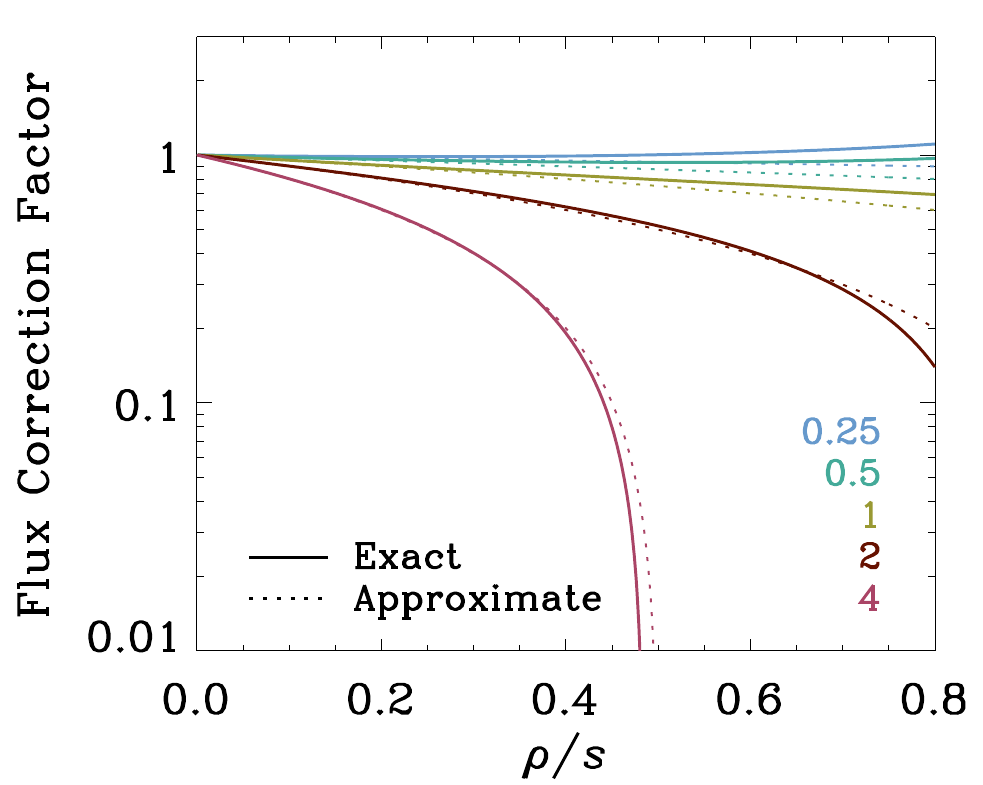}
\caption{
Flux correction factor as a function of $\rho / s$, the ratio between the aperture radius and the angular separation distance of the two partially overlapping comae 1 and 2, versus different measured flux ratios of $F'_2 / F'_1$ (colour coded as shown in the legend). The exact results and the approximate counterparts from Taylor expansion keeping the lowest order of $\rho / s$ are plotted as solid and dotted lines, respectively. At small $\rho / s$, the two results are largely inseparable, thereby validating the approximation.
\label{fig:fcorr}
} 
\end{center} 
\end{figure}

To obtain an approximation, we can employ a Taylor series expansion at $\rho/s = 0$ for $\mathcal{J}$:
\begin{equation}
\mathcal{J} = \frac{\pi}{2}\left(\frac{\rho}{s}\right)^2 \left[1 + \frac{1}{8}\left(\frac{\rho}{s}\right)^2 + \frac{3}{64}\left(\frac{\rho}{s}\right)^4 + \mathcal{O}\left(\left(\frac{\rho}{s}\right)^{6}\right)\right]
\label{eq_J2}.
\end{equation}
\noindent If the accuracy of keeping the lowest-order term of $\rho / s$ is deemed sufficient, which corresponds to $\rho \ll s$, Equations (\ref{eq_Freal}) and (\ref{eq_Fcor2}) will then become
\begin{equation}
F_1 \approx F'_1 - \frac{\rho}{2s}F'_2
\label{eq_Freal_approx}
\end{equation}
\noindent and
\begin{equation}
\frac{F_1}{F_1'} \approx 1 - \left(\frac{\rho}{2s}\right)\frac{F'_2}{F'_1}
\label{eq_Fcor3},
\end{equation}
\noindent respectively. The physical interpretation for the approximation is that the contaminating coma has a nearly uniform surface brightness throughout the region encompassed by the aperture. We show the comparison between the exact and approximate results with selected measured flux ratios $F'_2 / F'_1$ in Figure \ref{fig:fcorr}, where one can see that the approximation yields results largely indistinguishable from the exact ones at small $\rho / s$.

In cases where both comae have different surface brightness profiles but are still azimuthally symmetric, the actual flux of coma 1 is related to the measured fluxes of comae 1 and 2 by
\begin{equation}
F_1 \approx F'_1 - \frac{2 - n_2}{2} \left(\frac{\rho}{s}\right)^{n_2} F'_2
\label{eq_Freal_aprx_n},
\end{equation}
\noindent and the flux correction factor is
\begin{equation}
\frac{F_1}{F'_1} \approx 1 - \left[\frac{2 - n_2}{2} \left(\frac{\rho}{s}\right)^{n_2}\right] \frac{F'_2}{F'_1}
\label{eq_Fcor_aprx_n},
\end{equation}
where $n_2$ is the power-law index slope of the surface brightness of coma 2, assumed to be constant, and typically in a range of $1 \la n_j \la 2$ \citep{1987ApJ...317..992J}.

\restartappendixnumbering

\end{document}